\documentclass{aastex62}

\usepackage{mathtools}
\usepackage{epsfig}
\usepackage{comment}

\renewcommand{\thefootnote}{\fnsymbol{footnote}}

\received{April 4, 2018}
\revised{May 6, 2018}
\accepted{May 7, 2018}
\submitjournal{ApJ}

\shorttitle{Fermi analysis \& hadronic modeling of W28}
\shortauthors{Cui et al.}

\begin{document}

\title{Leaked GeV CRs from a broken shell: Explaining 9 years Fermi-LAT data of SNR W28}



\correspondingauthor{Yudong Cui, Paul K.H. Yeung}
\email{cuiyd@mail.sysu.edu.cn, paul2012@connect.hku.hk}

\author{Yudong Cui}
\affil{Institute of Physics and Astronomy, Sun Yat-Sen University, Zhuhai, 519082, China}

\author{Paul K.H. Yeung}
\affiliation{Institute of Astronomy and Department of Physics, National Tsing Hua University, Hsinchu, Taiwan}
\affiliation{Institute of Experimental Physics, Department of Physics, University of
Hamburg, Luruper Chaussee 149, D-22761 Hamburg, Germany}

\author{P.H. Thomas Tam}
\affiliation{Institute of Physics and Astronomy, Sun Yat-Sen University, Zhuhai, 519082, China}

\author{Gerd P\"uhlhofer}
\affiliation{Institut f{\"u}r Astronomie und Astrophysik, Eberhard Karls Universit{\"a}t T{\"u}bingen, Sand 1, D 72076 T{\"u}bingen, Germany}

\begin{abstract}
Supernova remnant (SNR)\,W28 is well known for its classic hadronic scenario, in which the TeV cosmic rays (CRs) released at early stage of this mid-aged SNR are illuminating nearby molecular clouds (MCs). Overwhelming evidences have shown that the northeast of the SNR (W28-North) has already encountered with the MC clumps. Through this broken shell -- W28-North, we believe the CRs with energy down to $<$1\,GeV to be able to be injected into nearby MCs. 
To further testify this hadronic scenario, we first analyse the 9 years Fermi-LAT data in/around W28 with energy down to 0.3\,GeV. Our Fermi-LAT analysis display a 10-200 GeV skymap which spatially matches well with the known TeV sources -- HESS\,J1801-233 (W28-North), HESS\,J1800-240\,A,\,B\,\&\,C (240\,A\,B\,\&\,C). At low energy band, we has discovered a 0.5-1\,GeV blob located to the south of 240\,B\,\&\,C, and a low flux of 0.3-1\,GeV at 240\,A. A hadronic model is build to explain our analysis results and previous multi-wavelength observations of W28. Our model consists of three CR sources: The run-away CRs escaped from a strong shock; The leaked GeV CRs from the broken shell -- W28-North; And the local CR sea. Through modelling the SNR evolution, CR acceleration\,\&\,releasing, we have explained the GeV-TeV emission in/around SNR\,W28 (except for 240\,A) in one model. Both the damping of the magnetic waves by the neutrals and the decreased acceleration efficiency are taken into account in our model due to the mid-age of SNR W28.

\end{abstract}



\section{Introduction} \label{sec:intro} 

For mid-aged/old supernova remnants (SNRs), the released TeV cosmic rays (CRs) mostly come from the run-away CRs during the early stage of the SNR, and they are expected filling the nearby environment almost homogeneously at present time. This hypothesis has been well testified in the case of SNR\,W28 \citep{Ah2008}, the H.E.S.S. data along with the NANTEN $^{12}$CO data has shown that the TeV sources -- HESS\,J1801-233 (W28-North) and HESS\,J1800-240 A, B, C (240\,A, 240\,B, 240\,C) spatially match well with their molecular cloud (MC) counterparts -- a MC at Northeast of the SNR (MC-N) and MCs to the South of the SNR (MC-A, MC-B, and MC-C). The following Fermi observations of SNR\,W28 by \cite{Abdo2010} has found GeV counterparts of W28-North and 240\,B, further Fermi analysis work by \cite{Hanabata2014} provided more detailed GeV spectrum of the southern counterparts, both their work suggest that this mid-aged SNR is likely releasing its GeV CRs into nearby MCs. Considering the relative lower diffusion coefficient of GeV CRs and their relative later releasing time when compared to the TeV CRs, it is not surprising to find that only some of these MCs (possibly due to the shorter distance to the SNR) are brightly illuminated in the Fermi-LAT skymap of \cite{Abdo2010}.

From the multi-wavelength view of W28, VLA 90cm radio image from \cite{Brogan2006} has shown a clear shell structure, which provides us the location (RA 18h01m42.2s, Dec -23$^\circ$20'6.0'') and the radius (0.34$^\circ$) of SNR\,W28. ROSAT/ASCA study by \cite{Rho2002} has discovered the thermal X-ray emission at the SNR center, and the estimated X-ray emitting mass is only $20-25 \,\rm{M_\odot}$. The lack of nonthermal X-ray emission suggests that this mid-aged SNR can no longer accelerate fresh super-TeV electrons. Interestingly, there is a bright feature at northeast of the SNR which shines in both radio and thermal X-ray. This bright feature which spatially coincides with W28-North indicates that SNR\,W28 has already encountered with nearby MCs, i.e. the MC-N, and shocked the gas there \citep{Rho2002,Nakamura2014,Zhou2014}. This shock-MC encounter scenario is also supported by other observational evidences: The northeast region of SNR\,W28 contains a rich concentration of 1720-MHz OH masers \citep{Frail1994,Claussen1999,Hewitt2009} (with VLSR in the range of $5-15 \mathrm{km\,s^{-1}}$), and near-IR rovibrational H$_2$ emission \citep{Reach2000,Neufeld2007,Marquez2010}; The velocity dispersion distribution of the NH$_3$ emission line at this NE region suggests an external disruption from the W28 SNR direction \citep{Nicholas2011,Nicholas2012}. More multi-wavelength observations of W28 can be found in section~\ref{sec:Constrains}.

So far, the releasing process of the GeV CRs from mid-aged SNRs is still an open question. In previous works on W28, see e.g. \cite{Gabici2010,Li2010,Ohira2011,Tang2017}, run-away CRs with energy down to $\lesssim1$\,GeV are able to escape the shock during the SNR evolution history. \cite{Ohira2011} has studied the case in which a fast shock is sweeping and passing small clumps and neither the dynamical evolution of the system nor the shock environments to confine CRs is affected by the small clumps. Additional to that, \cite{Ohira2011} has also explored a scenario in which the entire SNR shell is embedded inside MC clumps, and the GeV CRs used to be trapped at the acceleration region become the run-away CRs immediately after the shock-MC encounter. For SNR W28, clearly it has only partially collided with MC-N, and the shock can not simply "pass" MC-N unchanged, due to the big size of MC-N. 

In our work, we explore a model that most of the escaped $\lesssim$10GeV CRs come from the broken shell at W28-North. When parts of the SNR -- W28-North encounters with dense MC clumps -- MC-N, the shock is significantly slowed/stalled (see e.g. the stalled shock by dense clumps in case of RX\,J1713.7-3946 \citep{Sano2010,Gabici2016}) and the magnetic turbulence in both the upstream and downstream of the shock will be damped by the high density neutrals. This slowed/stalled shock together with the damping of magnetic turbulence allow all the GeV CRs at the acceleration region to escape (see e.g. \cite{Ohira2011}), and also allow the CRs behind the shock to be carried/diffusing into MC-N. Noticeably, through the MHD simulation of the shock-MC encounter, \cite{Inoue2012} suggested a shell of amplified magnetic turbulence at the location of the stalled shock, this shell would prevent GeV CRs from entering the MC clumps. However, this shell from \cite{Inoue2012} requires a fast shock ($2500\,\mathrm{km\,s^{-1}}$) hitting the MC clump and it can only last for a relative short time of $\lesssim 10^3\,$year. For the mid-aged SNR W28, the leakage of GeV CRs at W28-North is also supported by the millimetre observations by \cite{Vaupre2014,Maxted2016}: Ionization evidences have been found only at MC-N, very likely owing to the leaked 0.1-1GeV CRs from the broken shell. A detailed model of the leaking process at W28-North is described in section~\ref{sec:Leaked}. Additionally, we do not rule out that the run-away CRs from the entire SNR (most parts of the SNR have not encountered MC clumps yet, and they display a shock velocity of $\sim100\,\mathrm{km\,s^{-1}}$) are able to reach a very low energy of 1\,GeV and contribute significantly to the $\lesssim1\,$GeV $\gamma-$ray emission at MC-N, more discussions about this hypothesis can be found in section~\ref{sec:1GeV}.

Triggered by the idea that GeV CRs leaked from W28-North is probably dominating the GeV emission in/around W28, in the first part of our work (chapter~\ref{sec:Data}), using 9 years Fermi-LAT data, we re-analysis the previously discovered GeV sources in/around W28 with spectral energies down to 0.3\,GeV; in the second part (chapter~\ref{sec:Models}), we deliver a hadronic explanation for the GeV-TeV emission in/around W28 involving the leaked CRs from the broken shell -- W28-North.

\section{Fermi-LAT data Analysis} \label{sec:Data}

\subsection{Data preparation} \label{sec:data}
We performed a series of binned maximum-likelihood analyses for a 20$^\circ$$\times$20$^\circ$ region-of-interest (ROI) centered at RA=$18^{h}00^{m}30.000^{s}$, Dec=$-23^\circ26'00.00"$ (J2000), which is approximately the radio position of W28. We  used the data obtained with LAT between 2008 August 4 and 2017 April 4. The data were reduced and analyzed with the aid of \emph{Fermi} Science Tools v10r0p5 package. In view of the complicated environment of the Galactic plane regions, we adopted the events classified as Pass8 ``Clean" class for the analysis so as to better suppress the background. The corresponding instrument response function (IRF) ``P8R2$_-$CLEAN$_-$V6" is used throughout the investigation. We further filtered the data by accepting only the good time intervals where the ROI was observed at a zenith angle less than 90$^\circ$ so as to reduce the contamination from the albedo of Earth.

For subtracting the background contribution, we  included the Galactic diffuse background (gll$_-$iem$_-$v06.fits), the isotropic background (iso$_-$P8R2$_-$CLEAN$_-$V6$_-$PSF3$_-$v06.txt for ``PSF3" data, iso$_-$P8R2$_-$CLEAN$_-$V6$_-$FRONT$_-$v06.txt for ``FRONT" data or iso$_-$P8R2$_-$CLEAN$_-$V6$_-$v06.txt for a full set of data) as well as all other  point sources cataloged in the most updated Fermi-LAT catalog \citep[3FGL;][]{Acero2015a} within 25$^\circ$ from the ROI center in the source model.  Based on the $>$10 GeV morphological studies presented in  \S\ref{high}, we refined the source configuration in the W28 complex, which is spatially associated with HESS J1801-233 and HESS\,J1800-240\,A\,B\,\&\,C. We  set free the spectral parameters of the  sources within 6$^\circ$ from the ROI center in the analysis. For the sources beyond 6$^\circ$ from the ROI center, their spectral parameters were fixed at the catalog values. 

In spectral and temporal analysis, we required each energy-bin and time-segment to attain a signal-to-noise ratio $\gtrsim$$3.0\sigma$  (equivalently, a TS value $\gtrsim$9 and a chance probability $\lesssim$0.3\%) for a robust result. For each energy-bin or time-segment \emph{dissatisfying} this requirement, we placed  a 2.5$\sigma$ upper limit on its flux.

\subsection{Morphological Analysis} \label{sec:spatial}

\subsubsection{$>$10 GeV} \label{high}

\begin{figure}
	\centering
	\includegraphics[width=0.95\linewidth]{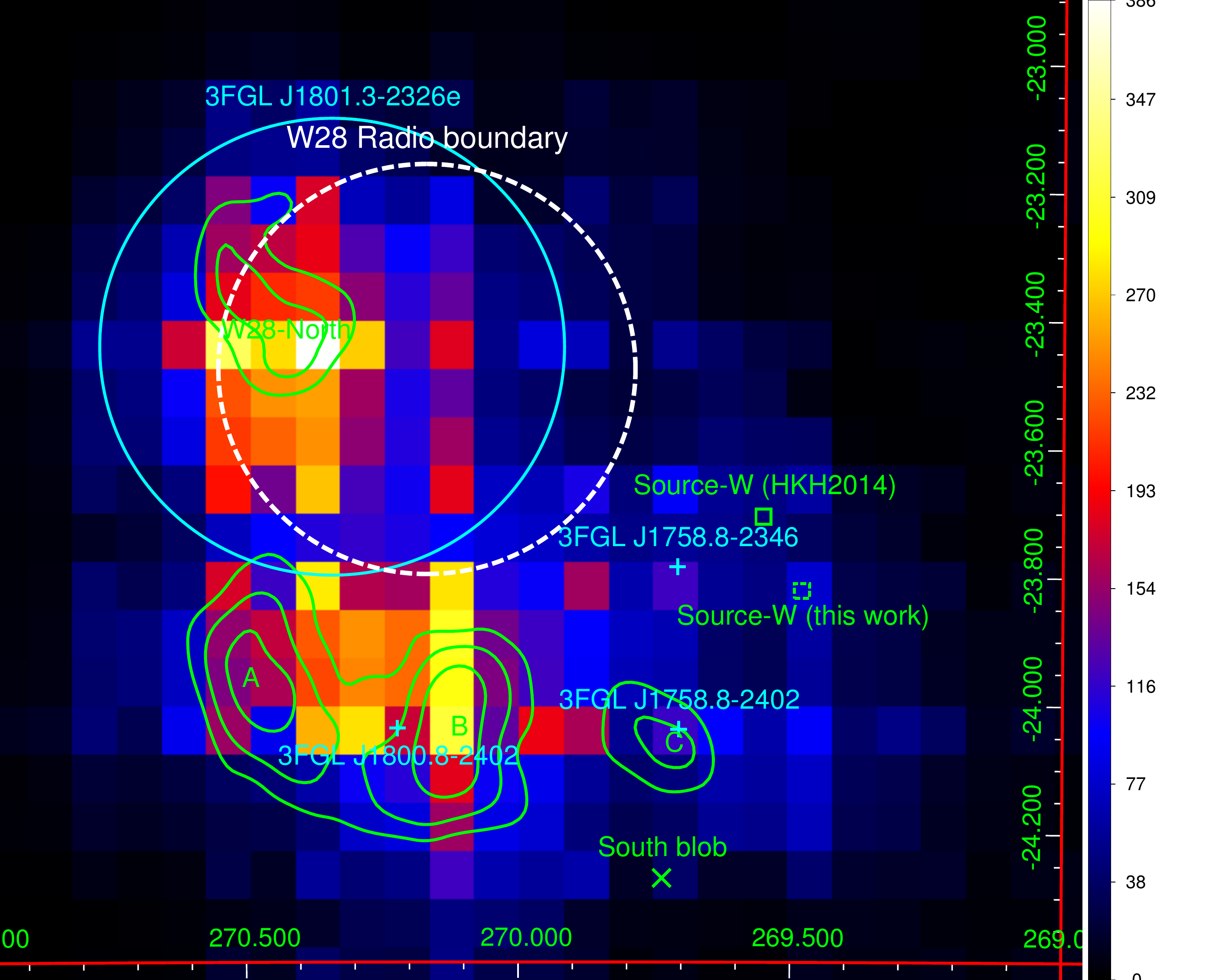}
	\caption{TS map of the entire W28 region at 10-200\,GeV, with Right ascension/Declination as x/y axis. The radio boundary of SNR\,W28 is marked in white dashed circle. The H.E.S.S. morphology contour is presented in green lines ($4,5,6\sigma$), and all the TeV sources HESS\,J1801-233 (W28-North) and HESS\,J1800-240 A, B, C (A,B,C) are indicated in green as well. The 3FGL catalog sources are marked in cyan circle and crosses. The new discovered GeV sources -- Source-W and South blob are marked in green boxes and cross, here Source-W is first discovered by \cite{Hanabata2014} and its position by \cite{Hanabata2014} is marked in a solid green box with a note -- ``HKJ2014". More details about Source-W and South blob can be found in Fig.~\ref{tsmap_small}~\&~\ref{tsmap_low} and the text. }
	\label{tsmap}
\end{figure}

We investigate the morphology of the W28 complex region in the 10$-$200 GeV regime. The test-statistic (TS) map of this field is shown in Figure~\ref{tsmap}, where all 3FGL catalog sources except 3FGL J1801.3-2326e (the northeastern part of W28), 3FGL J1800.8-2402, 3FGL J1758.8-2402 and 3FGL J1758.8-2346 are subtracted.   The peak detection significance is $\sim$$20\sigma$ and is in the northeastern part.  

In the source model, we proceeded to replace 3FGL J1800.8-2402 and 3FGL J1758.8-2402 with three point sources at the positions of 240\,A, B \& C respectively. In order to revise the morphology of 3FGL J1801.3-2326e  (originally a uniform disk with a $0.39^\circ$ radius in the northeastern part), we followed a scheme of a likelihood ratio test adopted by  \citet{Yeung2016} and \citet{Yeung2017a}. Since its centroid is exactly at the center of 3FGL J1801.3-2326e, we varied only the radius of extension while remaining the center unchanged.  We assigned it a simple power-law.  The $-ln(likelihood)$ of different radii in 10$-$200 GeV are tabulated in Table~\ref{Ext}, and the most likely radius is determined to be $0.345^\circ$. We therefore adopted this morphology for the northeastern part of W28 in subsequent analyses, and we refer to it as \emph{Fermi} J1801.4-2326 so as to avoid confusion with 3FGL J1801.3-2326e. 

\begin{figure}
	\centering
	\includegraphics[width=0.54\linewidth]{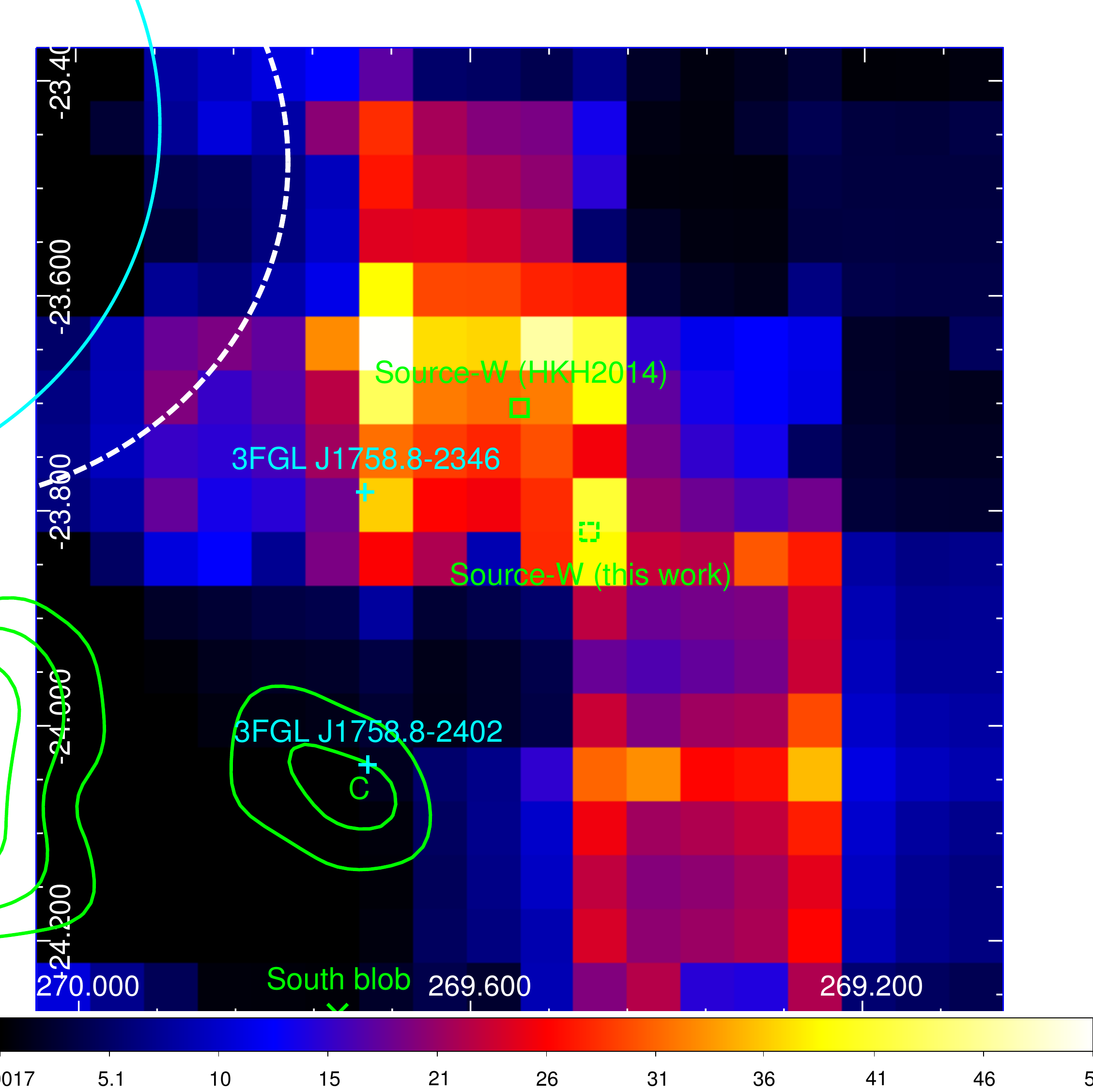}
	\caption{TS map of Source-W at 10-200\,GeV. Same marks described in Fig.~\ref{tsmap} are used here.}
	\label{tsmap_small}
\end{figure}

We further re-created the TS map with all sources in the revised model except 3FGL J1758.8-2346 (the western clump) subtracted, and it is presented in Figure~\ref{tsmap_small}. We thus revised the position of 3FGL J1758.8-2346 to be (269.47917$^\circ$, $-$23.820494$^\circ$)$_{J2000}$, which is the centroid determined on this map, and we re-named it ``Source-W". Here, we finalised the source model for the spectral analysis.

\subsubsection{$<$10 GeV} \label{low}

\begin{figure}
	\centering
		\includegraphics[width=0.48\linewidth]{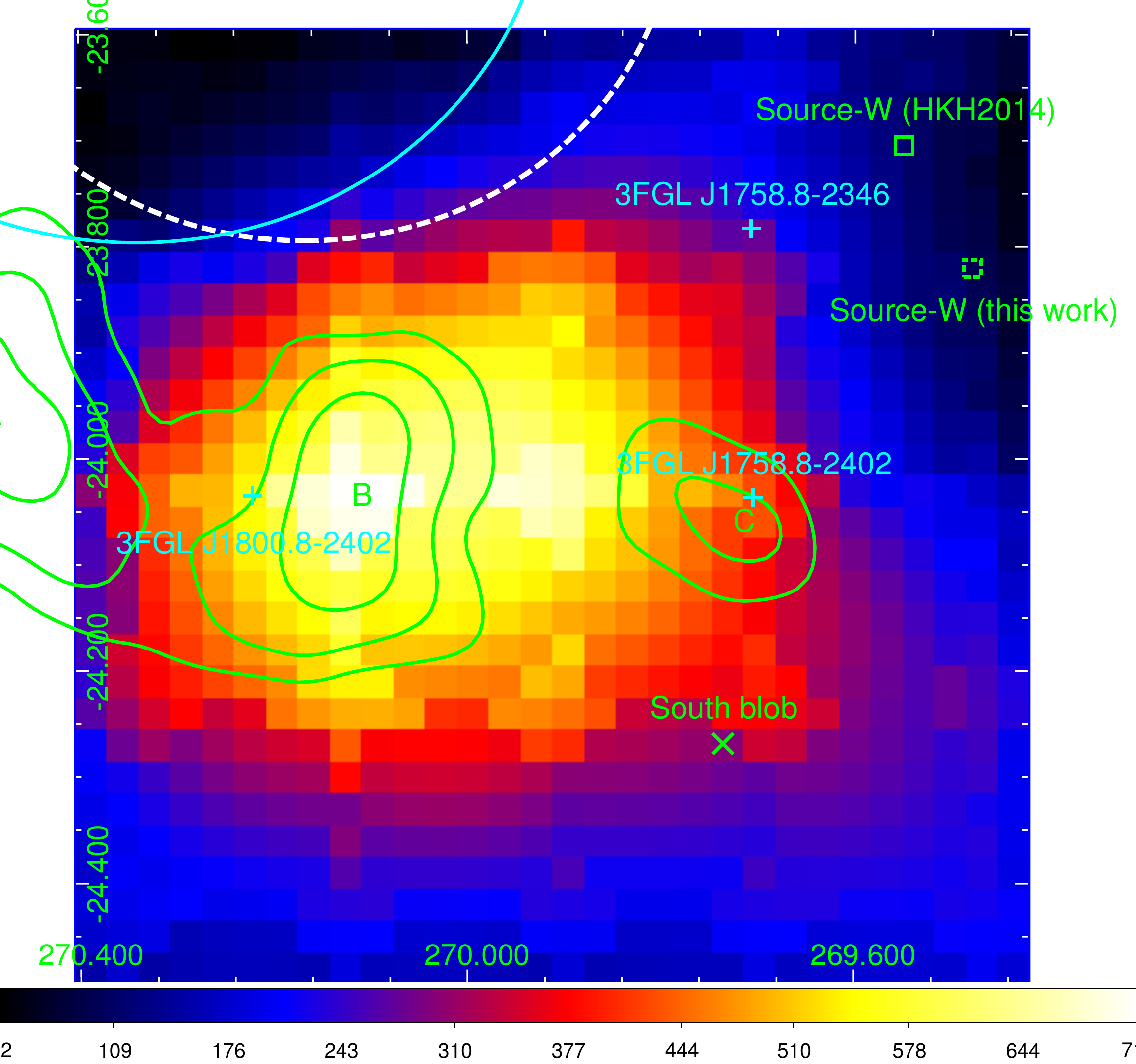}
		\includegraphics[width=0.48\linewidth]{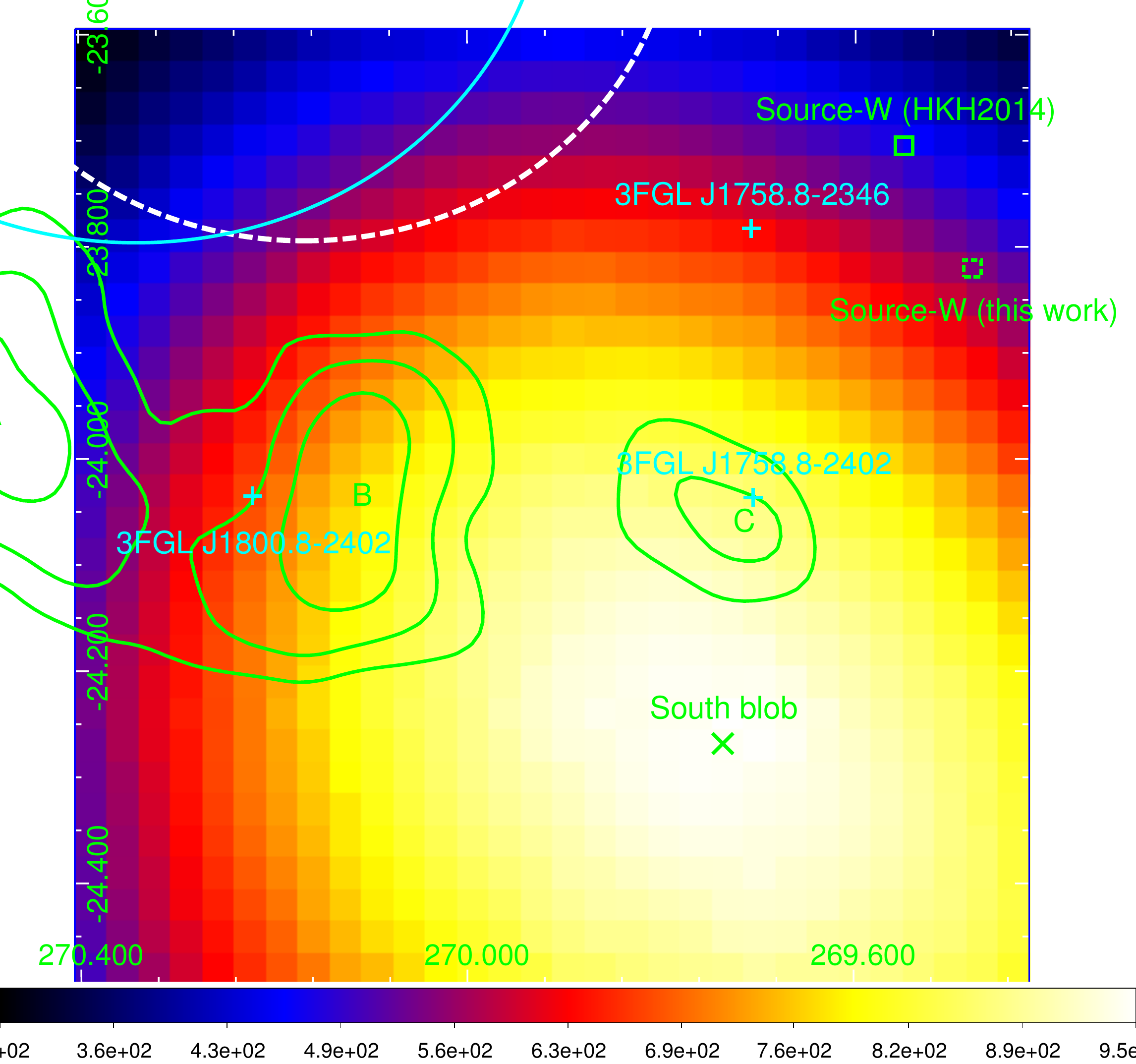}
	\caption{TS maps of South blob at 1-50\,GeV (left panel) and 0.5-1\,GeV (right panel). Same marks described in Fig.~\ref{tsmap} are used here.}
	\label{tsmap_low}
\end{figure}

Motivated by the LAT spectra of 240\,B \& C presented in \S\ref{sec:spectral}, we also investigate their morphology  in lower energy bands. We created TS maps where all sources except 240\,B \& C are subtracted (Figure~\ref{tsmap_low}), in 1-50 GeV (left) and 0.5-1 GeV (right) respectively. We adopted ``PSF3" data for optimisation of spatial resolution.

In 1-50 GeV, the excess shown on the map is coincident with 240\,B \& C, but the excesses from these two spatial components are hardly resolved. In 0.5-1 GeV, the excesses from 240\,B \& C, if any, are buried under the PSF wing of a brighter blob (south blob) whose centroid is at (269.73584$^\circ$, $-$24.268803$^\circ$)$_{J2000}$. The detection significance of this south blob is $\gtrsim30 \sigma$.

\begin{figure}
	\centering
		\includegraphics[width=0.48\linewidth]{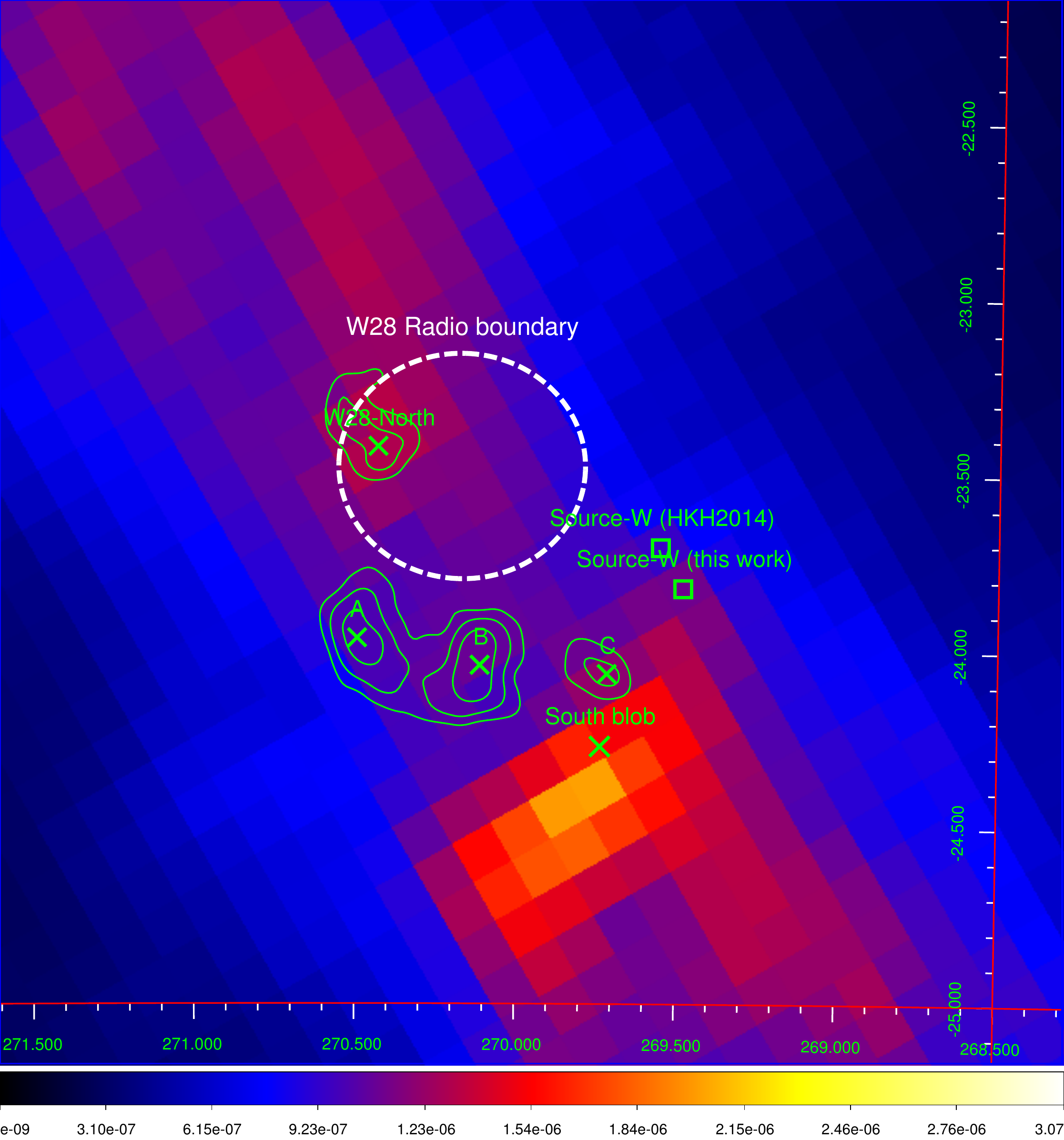}
	\caption{Skymap of diffuse background at 0.7GeV drawn from the file of gll$_-$iem$_-$v06.fits. The unit here is $\mathrm{photon\,s^{-1}\,sr^{-1}\,MeV^{-1}\,cm^{-2}}$. Same marks described in Fig.~\ref{tsmap} are used here.}
	\label{fig:background}
\end{figure}

The diffuse background which is mainly caused by the CR sea interacting with interstellar medium are particularly important at $<$10 GeV band. As seen in Fig.~\ref{fig:background}, the south blob is located inside a bright background region, therefore it is also possible that the south blob is caused by an unclean background reduction.

\subsection{Spectral Analysis} \label{sec:spectral}

To construct the binned spectra of our five targeted sources, we performed an independent fitting of each spectral bin.  For each spectral bin, we assigned  PL models to all targeted sources. Considering that we include photons with energies below 1 GeV, and that we are investigating crowded regions in the Galactic plane, we find it \emph{inappropriate} to adopt a full set of data whose large PSF (e.g., a 68\% containment radius of $\sim$2.3$^\circ$ at 0.3 GeV; cf. SLAC\footnote[2]{\label{slac}Fermi LAT Performance: \url{http://www.slac.stanford.edu/exp/glast/groups/canda/lat_Performance.htm}}) leads to severe source confusion. Meanwhile, adopting only ``PSF3" data is also \emph{discouraged} in spectral fittings because of large systematic uncertainties induced by severe energy dispersion. For ``FRONT" data in 0.3$-$1 GeV, the FWHM of its PSF is $<$75\% of that for a full set of data and its energy dispersion is greater than that for a full set of data by only $\lesssim$1\% of the photon energy (cf. SLAC~$^{\ref{slac}}$). Therefore, we adopted only ``FRONT" data in order to achieve a compromise between a small PSF and lessened energy dispersion.

\begin{figure}
	\centering
		\includegraphics[width=0.98\linewidth]{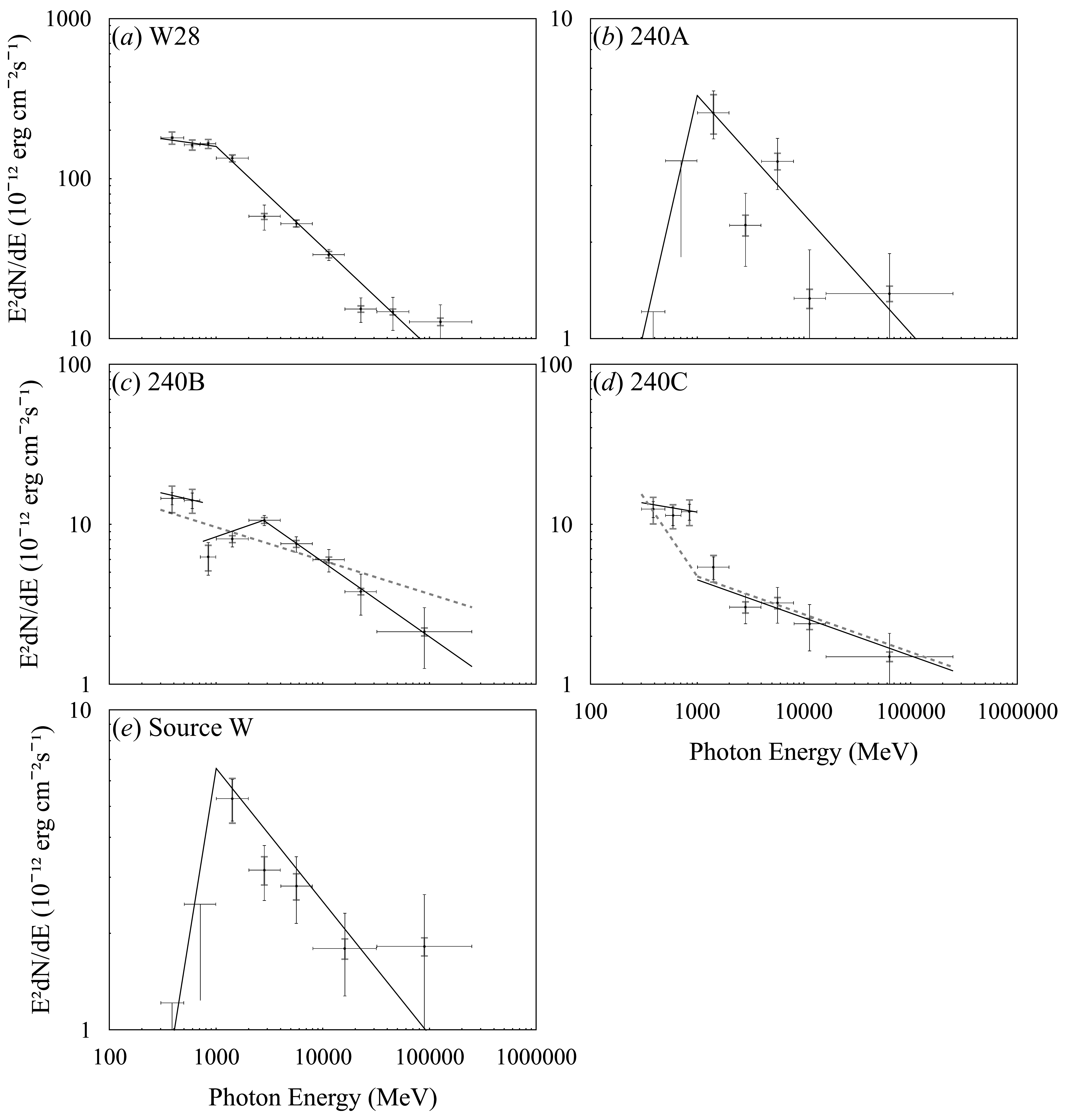}
	\caption{0.3-250 GeV SEDs. On panels (a), (b) and (e), the solid lines indicate the best-fit BKPL models. On panels (c) and (d), solid lines demonstrate models with discontinuous flux, while gray dashed lines demonstrate models with continuous flux. Statistic uncertainties are plotted in black while systematic uncertainties are plotted in grey. The systematic uncertainties consist of two parts:
(i) The differences when varying the Galactic diffuse emission by $\pm$5\%.
(ii) \href{https://fermi.gsfc.nasa.gov/ssc/data/analysis/scitools/Aeff_Systematics.html}{The uncertainties on LAT effective area.}}
	\label{SED}
\end{figure}

The 0.3-250 GeV spectral energy distributions (SEDs)  are shown in Figure~\ref{SED}.  We examined how well each spectrum  can be described by, respectively, a simple power-law (PL)
\begin{center}
	$\frac{dN}{dE}=n_0(\frac{E}{E_0})^{-\Gamma}$ \ \ \ ,
\end{center}
and a broken power-law (BKPL)
\begin{center}
	$\frac{dN}{dE}=\begin{cases} n_0(\frac{E}{E_\mathrm{b}})^{-\Gamma_1} & \mbox{if } E<E_\mathrm{b} \\ n_0(\frac{E}{E_\mathrm{b}})^{-\Gamma_2} & \mbox{otherwise} \end{cases}$ \ \ \ .
\end{center}
For BKPL, we fixed the spectral break at $E_\mathrm{b}=1$ GeV because it cannot be properly constrained. The results of spectral fitting are tabulated in  Table~\ref{spectral1}.

For  \emph{Fermi} J1801.4-2326, the likelihood ratio test indicates that  BKPL is preferred over PL by $>$$13\sigma$.  A BKPL model yields a photon index $\Gamma_{1}=2.10 \pm 0.03$ below the spectral break $E_\mathrm{b}=1$ GeV and a photon index $\Gamma_{2}=2.63 \pm 0.02$ above the break.  In view of the apparent bump above 30 GeV, we repeated this test with excluding the 0.3$-$1 GeV data. It turns out that the 1-250 GeV spectrum of \emph{Fermi} J1801.4-2326 is satisfactorily described by PL with $\Gamma=2.64 \pm 0.02$, while the additional parameters of BKPL are \emph{not} strongly required (only $\sim$$1.5\sigma$). The fitting parameters are tabulated in Table~\ref{spectral2}.

For the GeV counterpart of 240\,A,  BKPL is preferred over PL by $\sim$$3.3\sigma$.  A BKPL model yields a photon index $\Gamma_{1}=0.53 \pm 0.72$ below 1 GeV and a photon index $\Gamma_{2}=2.37 \pm 0.09$ above 1 GeV. 
For Source-W, BKPL is preferred over PL by $\sim$$4.3\sigma$.  In its BKPL model, the photon indices below and above 1 GeV are $\Gamma_{1}=-0.09 \pm 0.85$ and $\Gamma_{2}=2.42 \pm 0.10$ respectively.

For 240\,B \& C, each of their LAT spectra appears to contain a discontinuity of flux, resulting that both PL and BKPL models fail to describe their spectral shapes (i.e., each of them might be decomposed into two spectral components). With regards to this, we looked into each of their spectral shapes in two mutually exclusive energy bands decoupled at around the termination point of the first component (0.75 GeV and 1 GeV for 240\,B \& C respectively). The results of spectral fitting are tabulated in  Table~\ref{spectral2}.

It turns out that, in 0.3-250 GeV, a two-component scenario (a model with discontinuous flux) is preferred over a single-component scenario (a model with continuous flux) by $>$$10\sigma$ for both 240\,B \& C. For each source, we compared the sum of PL TS values in the two decoupled energy bands with the BKPL TS value in 0.3-250 GeV. Since a two-independent-PL model can be reduced to a BKPL model by simply uniting the prefactors, the number of additional d.o.f. is only 1.
The first LAT component of 240\,B has a photon index $\Gamma=2.15 \pm 0.39$. Its second LAT component starts with $\Gamma_{1}=1.77 \pm 0.15$ at 0.75 GeV, and then softens (at a $\sim$$3.2\sigma$ significance) to $\Gamma_{2}=2.47 \pm 0.09$ above a break energy $E_\mathrm{b}=2780 \pm 837$ MeV.
240\,C has its first LAT component with $\Gamma=2.11 \pm 0.21$, and its second LAT component starts at 1 GeV with a photon index $\Gamma=2.24 \pm 0.10$. A spectral break is not required at all for its second component.
Probably, the 0.3-1 GeV emissions detected at 240\,B \& C are mostly originated from a south blob (see Figure~\ref{tsmap_low}). With regards to this, the  0.3-1\,GeV data points of 240\,B \& C can only serve as upper-limits in our final results. 

\begin{figure}
	\centering
		\includegraphics[width=0.98\linewidth]{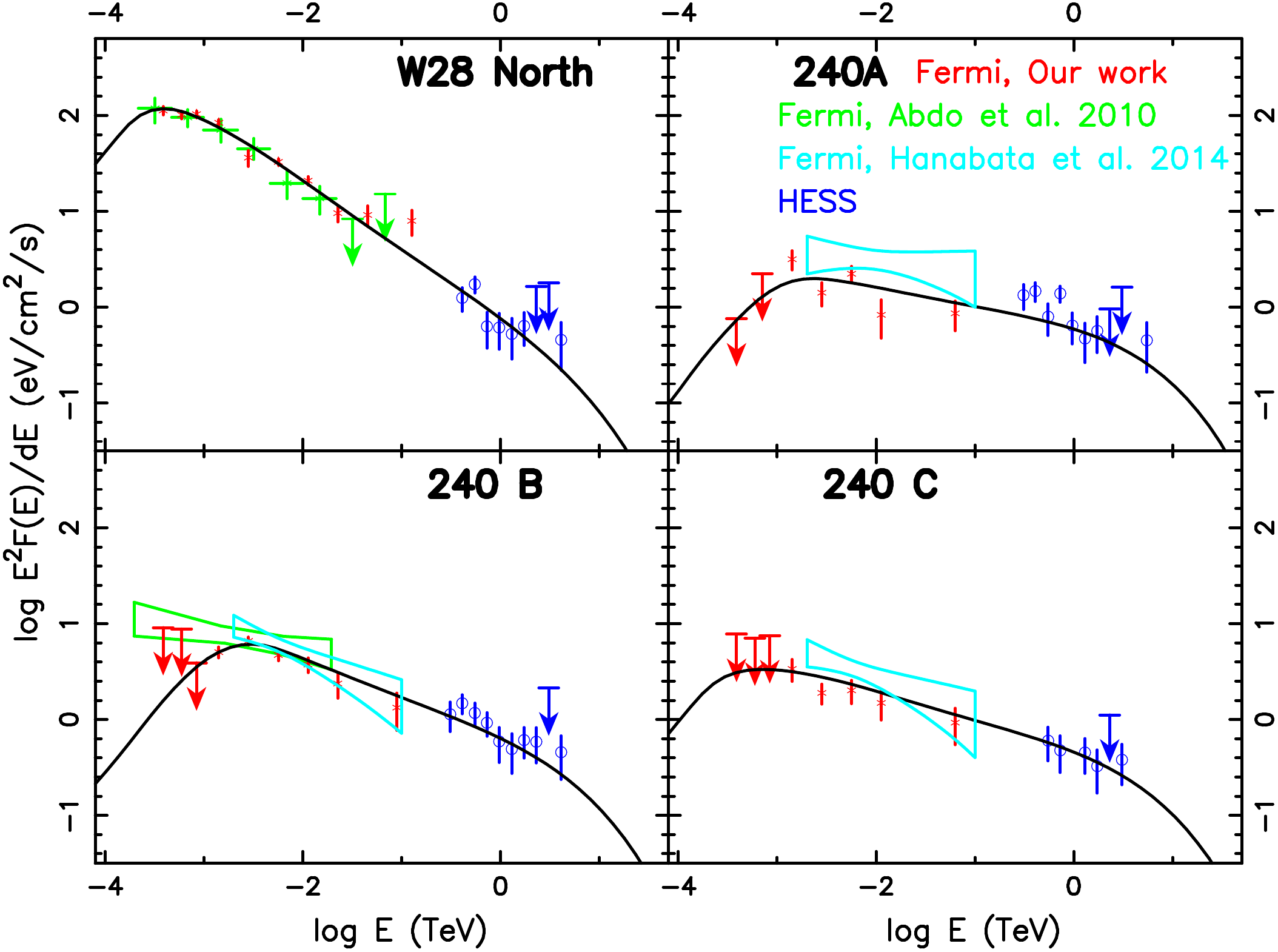}
	\caption{Observational $\gamma$-ray spectra of W28. Fermi-LAT observations in/around W28 from our work and previous work \citep{Abdo2010,Hanabata2014} are shown in red, green, and cyan, respectively. The HESS data points from \cite{Ah2008} is marked in blue. Assuming a simple power-law CR population at each target MC clumps, the pre-model fitting results are shown in black lines.}
	\label{fig:pre_model}
\end{figure}

When compared with previous Fermi-LAT analysis work, as seen in Fig.~\ref{fig:pre_model}, we deliver similar spectral results of the relative brighter GeV sources -- W28-North and 240\,B, while our GeV fluxes of 240\,A,\&\,C are lower than the ones from \cite{Hanabata2014}. In Fig~\ref{fig:pre_model}, we also plot a simple hadronic fitting (with a power-law CR population) for each $\gamma$-ray source. The CR power-law indices in the fitting of W28-North, 240\,A,\,B,\,\&\,C are set as 2.7, 2.2, 2.4, \& 2.3, respectively. Due to the low $<1$GeV emission at 240\,A\,\&\,B, minimum cutoffs of the CR population of 5\,GeV \& 9\,GeV are used in the fitting of 240\,A\,\&\,B, respectively. 

Source-W is ignored in this simple fitting and in the hadronic model below, because it has neither distinct MC counterparts nor TeV counterparts (it does have a radio counterpart). Here we do not exclude the possibility that Source-W could be explained by CRs released from the SNR as well, e.g., \cite{Hanabata2014} has presented a successful hadronic model for Source-W, in which they have used a mass of the MC at Source-W of $3.5\times10^3\,$M$_\odot$ and a distance of $16\sim25\,$pc to the SNR. However, under one SNR model and one diffusion coefficient, it seems difficult to explain why the GeV flux from Source-W is almost comparable to the ones from 240\,A,\,B,\,C, as seen in \cite{Hanabata2014}, in order to compensate the relative lower mass of the MC counterpart of Source-W, a much higher total CR energy is adopted in the Source-W model when compared to the ones adopted in their models of 240\,A,\,B,\,C. More complex hadronic scenarios could help solving this problem, e.g., by introducing a magnetic tube connecting the SNR and Source-W; by introducing an asymmetric SNR expansion, that the shock can carry GeV CRs all the way to Source-W, as discussed by \cite{Hanabata2014}, this scenario can also explains the extended radio structure at Southwest (Source-W) of the SNR. Additionally, \cite{Hanabata2014} has also discussed the other possible explanations (pulsar and blazar) for Source-W. 

\section{The hadronic model} \label{sec:Models}
\subsection{The physical constrains from multi-wavelength observations of W28}
\label{sec:Constrains}
Based on the observational data introduced above, SNR\,W28 is obviously an mid-aged SNR and can no longer accelerate fresh super-TeV electrons, and it has already encountered with MC-N, part of the CRs trapped in/behind the shock are able to be released. Further constrains of our physical parameters by multi-wavelength observations are list below.
\begin{itemize}
\item {\it Distance to Earth}, $\sim2\,\mathrm{kpc}$ {\it and SNR radius,} $\sim13\,\mathrm{pc}$. SNR\,W28 is located inside a complex star-forming region, where H{\sc ii} region M\,20 (d $\sim$ 1.7\,kpc \citealt{Lynds1985}) and M\,8 (d $\sim$ 2\,kpc \citealt{Tothill2002}) are seen, nonetheless, there is no solid evidence to link these H{\sc ii} regions with SNR\,W28. \cite{Velazquez2002} suggested the distance of SNR\,W28 as $\sim1.9$\,kpc when associated it with a $70\mathrm{M_\odot}$ H{\sc i} feature detected at the SNR region, this H{\sc i} feature is also seen as the evidence of the interaction between SNR\,W28 and its surrounding gas.
\item {\it The density of pre-SN circumstellar medium}, $\sim5 \,\rm{cm^{-3}}$. Through the near-infrared and millimeter-wave observation, \cite{Reach2005} found that the different morphologies of W28 at different wavelengths can be explained by a highly nonuniform structure of giant molecular clouds, with low-density inter-clump medium (ICM) ($n_\mathrm{H}\sim 5 \,\rm{cm^{-3}}$) occupying most ($90\%$) of the volume, moderate-density clumps ($n_\mathrm{H}\sim 10^3 \,\rm{cm^{-3}}$) occupying most of the rest of the volume, and dense cores. \cite{Velazquez2002} has derived a H{\sc i} density upper-limits of $n_{\mathrm{HI}}\sim 1.5-2\,\rm{cm^{-3}}$, which is in accordance with the ICM density mentioned above, assuming that observed mass (swept H{\sc i} gas around the SNR) are evenly distributed inside a bubble with a 20\,pc radius. 
\item {\it Shock velocity at present time}, $\sim100\,\mathrm{km\,s^{-1}}$. Through observing the neutral hydrogen around the SNR, a H{\sc i} cloud is detected by \cite{Velazquez2002} near $V_\mathrm{LSR}=+37\,\mathrm{km\,s^{-1}}$, overlapping the center of W28. This expanding H{\sc i} cloud is likely to be a swept thick H{\sc i} shell surrounding the SNR, hence, the velocity dispersion of this H{\sc i} cloud could be lower than the intrinsic shock velocity. A more accurate method is through directly measuring the forbidden lines at the shock downstream, \cite{Bohigas1983} estimated shock velocities of W28 between 60\,km\,s$^{-1}$ and 90\,km\,s$^{-1}$ using the line strength ratio of O{\sc iii}\,$\lambda$5007/H$_\alpha$, while \cite{Long1991} derived velocities larger than 70\,km\,s$^{-1}$ from line strength ratio of N{\sc ii}/H$_\alpha$ and S{\sc ii}/H$_\alpha$. 
\item {\it The time of the shock-MC encounter at W28-North}. Detailed XMM study on W28 by \cite{Zhou2014} has found a soft components ($\sim0.3$\,keV, $\sim1\,\mathrm{M_\odot}$) and a hard components ($\sim0.6$\,keV, $\sim0.2\,\mathrm{M_\odot}$) at NE shell (W28-North) whose ionization timescales are estimated to be $>7.5$\,kyr and $10-40$\,kyr, respectively. In our model below, the shock-MC encounter time is set at 25\,kyr after the SN explosion (12\,kyr ago from present time).
\item {\it SN total Energy}, $\sim 1\,\mathcal{E}_{51}\,(10^{51}\mathrm{erg})$. This is the typical total energy for both the core collapse (CC) SNe and the type Ia SNe, and it also satisfy the requirements of the derived circumstellar gas density and the observed shock velocity at present time (see more details in the SNR evolution model below).
\item {\it Progenitor mass}, $8\,\mathrm{M_\odot}$. Massive stars who often end into core collapse (CC) SNe are likely to be born in clusters inside/near giant molecular clouds \citep{Smartt2009}. Nonetheless, no central compact object is found in SNR\,W28 so far, and this is not surprising when consider the cooling of a neutron star (if it is a CC SN) at an age of $\sim$40\,kyr \citep{Yakovlev2004}. The small X-ray emitting mass found in the center of the SNR could be due to the SNR expansion inside an empty pre-SN bubble or other processes, such like the evaporation \citep{Rho2002,Zhou2014}. In our model below, we adopt a CC SN scenario with a progenitor mass of $8\,\mathrm{M_\odot}$ and an ejecta mass of $6\,\mathrm{M_\odot}$. Discussions about other type SNRs can be found in section~\ref{sec:result}.
\item {\it Diffusion coefficient}. A diffusion coefficient of $10\%$ of the Galactic standard ($D(E)=10^{27}(E/10\,\rm{GeV})^\delta cm^2/s$, $\delta=0.5$) is adopted for the entire space outside the SNR. This value is mainly based on the TeV spectrum fitting of all sources in/around W28, as well as on the GeV spectrum fitting of W28-North, see more details in section~\ref{sec:diffusion}. Similar values are also adopted by \citet{Gabici2010} who has used an approximation of point CR source to explain the GeV-TeV emission in/around W28.

\item {\it The masses of MC-N, A, B} are estimated as $5\times10^4\,$M$_\odot$, $4\times10^4\,$M$_\odot$, and $6\times10^4$\,M$_\odot$, respectively, via NANTEN $^{12}$CO data \citep{Gabici2010,Ah2008}. Noticeably, most components of these MCs are found covering a broad velocity range from $10\,\mathrm{km\,s^{-1}}$ to $20\,\mathrm{km\,s^{-1}}$ which corresponds to a kinematic distance of approximately 2 to 4\,kpc \citep{Ah2008}. In section~\ref{sec:result}, we discuss a scenario in which the MC clumps are put at 4kpc to Earth rather than at 2kpc. The mass of MC-C is estimated as $1.4\times10^4\,$\,M$_\odot$ by \cite{Nicholas2012}, however, in our model below, we adopt $2\times10^4\,$\,M$_\odot$ as the mass of MC-C for a better fitting.  
\item {\it The projected distances of HESS\,J1800-240\,A, B, C} from the center of the SNR are $\sim$20\,\rm{pc}, assuming the distance of SNR\,W28 to Earth as 2\,kpc \citep{Ah2008}. 
\end{itemize}

 \subsection{Models}
\subsubsection{The SNR evolution}\label{sec:SNR}
For a core collapse SN with the minimum progenitor mass ($8\mathrm{M_\odot}$), the progenitor wind bubble can be neglected. Therefore, our SNR can directly expand into the ICM from the beginning. Our modelling results are only sensitive to the pre-SN environment (e.g., the progenitor wind bubble) but not to the ejecta mass, therefore, a type Ia SNR scenario would deliver similar results to the ones shown below.

\begin{figure}[!h]
   \centering
\begin{tabular}{c }
\includegraphics[width=13.25cm]{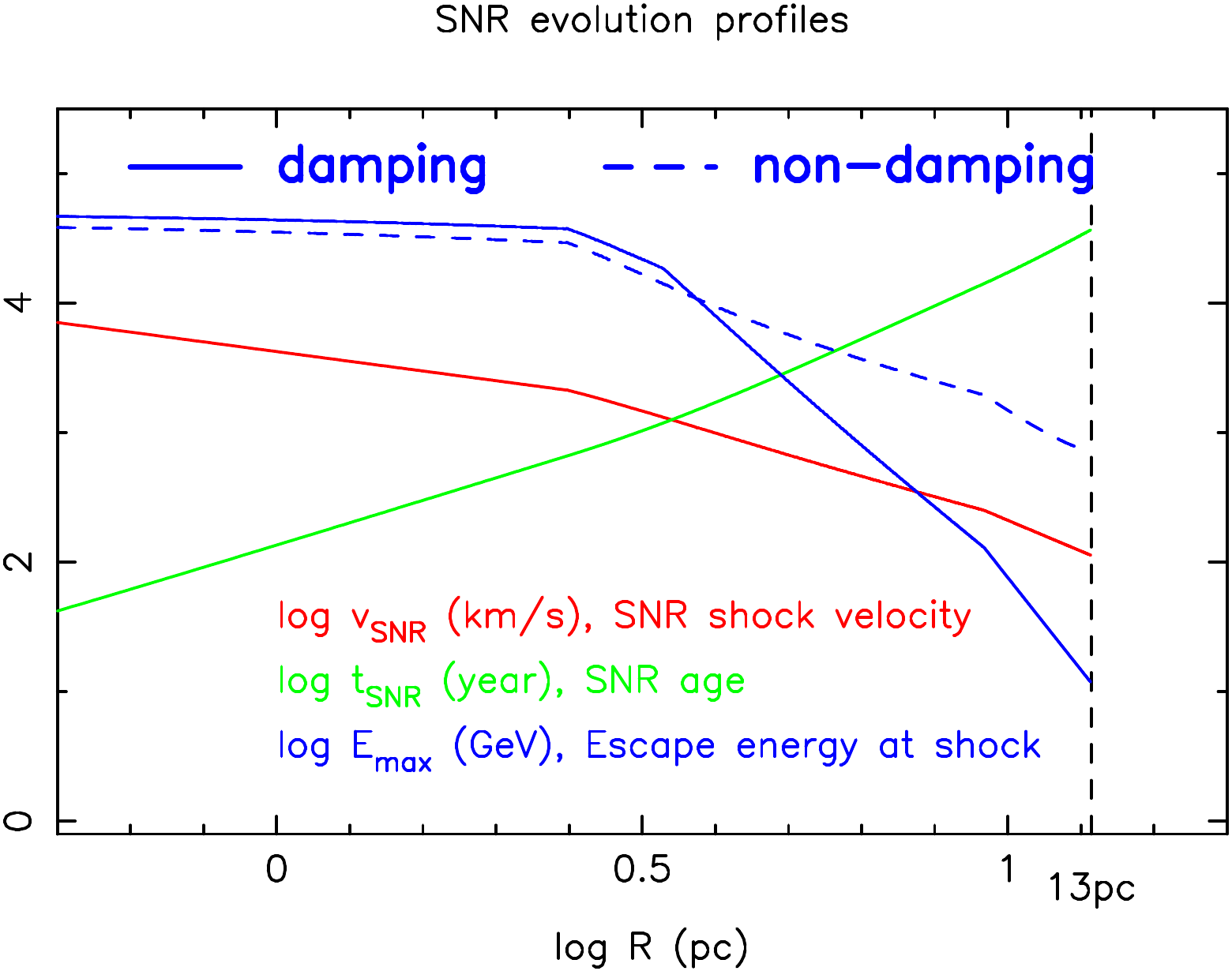} 
   \end{tabular}
   \vspace{-0.2cm}
   \caption{SNR evolution profiles. The same SNR age/velocity profile adopted in both the damping and non-damping models is shown in green/red line. Due to the different acceleration efficiencies and neutral densities adopted in the two models, two different evolution profiles of escape energy (blue solid/dashed lines) are presented. }
   \label{fig:SNR}
\end{figure}

In our work, we adopt the analytical solutions for the ejecta-dominated stage and the Sedov-Taylor stage from \cite{Chevalier1982,Nadezhin1985} and \cite{Ostriker1988,Bisnovatyi1995,Ptuskin2005}, respectively. When the shock is further slowed, we adopt the analytical solution for the pressure-driven snowplow (PDS) stage derived by \citet{Cioffi1988}. As shown in Fig.~\ref{fig:SNR}, inside a homogeneous ICM with density $n_\mathrm{ISM}=6\,\rm{H\,cm^{-3}}$, the SNR spends its first $\sim650$\,years ($\sim2.5$\,pc) in the ejecta-dominated stage, at $\sim14$\,kyr ($\sim9.3$\,pc) it finishes the Sedov-Taylor stage and enters the pressure-driven snowplow stage, when it finally reaches 13\,\rm{pc} at 37\,kyr, the present shock velocity is 110\,km\,s$^{-1}$. The escape energy of the run-away CRs, which is marked as blue lines in Fig.\ref{fig:SNR} is derived from the CR acceleration theory below.

\subsubsection{The CR acceleration and run-away CRs}
\label{sec:run-away} 
CRs acceleration in the SNR is well known as the first order Fermi acceleration, which normally provides a power-law CR spectrum at the shock with an index $\Gamma_\mathrm{CR}\sim-2.0$ and a cut off at the escape energy $E_\mathrm{max}$. 
To explain the super-TeV CRs observed in many young SNRs, \cite{Bell2004,Zirakashvili2008} have developed the acceleration theory of nonresonant streaming instability, in which the magnetic turbulence at the shock upstream is quickly amplified by the CR streaming, and is finally able to boost the escape energy up to hundreds of TeV in a young SNR. 

In a strong shock, only the CRs with energy above $E_\mathrm{max}$ are able to escape from the upstream and propagate to nearby environment with a flux $J_\mathrm{out}$, while other CRs which are trapped at the shock follow a power-law spectrum with an index $\Gamma_\mathrm{CR}\sim-2.0$. For a young SNR with a high shock velocity $v_\mathrm{SNR}\gtrsim 1000\,\mathrm{km\,s^{-1}}$, only those super-TeV CRs are expected to be released.
These early escaped super-TeV CRs could explain well the TeV emission from the MCs around mid-age SNRs, e.g., SNR\,W28 \citep{Gabici2010}, and even around young SNRs, e.g., SNR\,HESS\,J1731-347 \citep{Cui2016}.
With the Fermi-LAT observational study by \cite{Abdo2010}, the discovered GeV emission from W28-North which peaks at $\lesssim0.3$GeV indicates alternative sources of CRs. Sticking to the run-away CR explanation, one would need to gradually decrease the escape energy to $\lesssim1\,$GeV, mainly through decreasing the shock velocity, decreasing the magnetic field in the upstream, and/or increasing the neutrals density in the upstream. 

Once the shock velocity $v_{\mathrm{SNR}}$ and the density of the nearby circumstellar medium are obtained, the CR acceleration processes can be calculated through the acceleration theory of nonresonant CR streaming from \citet{Zirakashvili2008}. An analytical approximation of this theory derived by \cite{Zirakashvili2008} is adopted in our work, it provides the escape CR flux $J_\mathrm{out}$, the CR density at shock $n_0$, and most importantly -- the escape energy $E_\mathrm{max}$, see more details in \cite{Zirakashvili2008,Cui2016}. A magnetic field of $B_0=5\mathrm{\mu G}$ and an initial magnetic fluctuation (not amplified by the CR streaming yet) of $B_\mathrm{b}=7\%B_0$ in the ICM are used for the calculation of acceleration process. 

\begin{figure}[!h]
   \centering
\begin{tabular}{c }
\includegraphics[width=13.25cm]{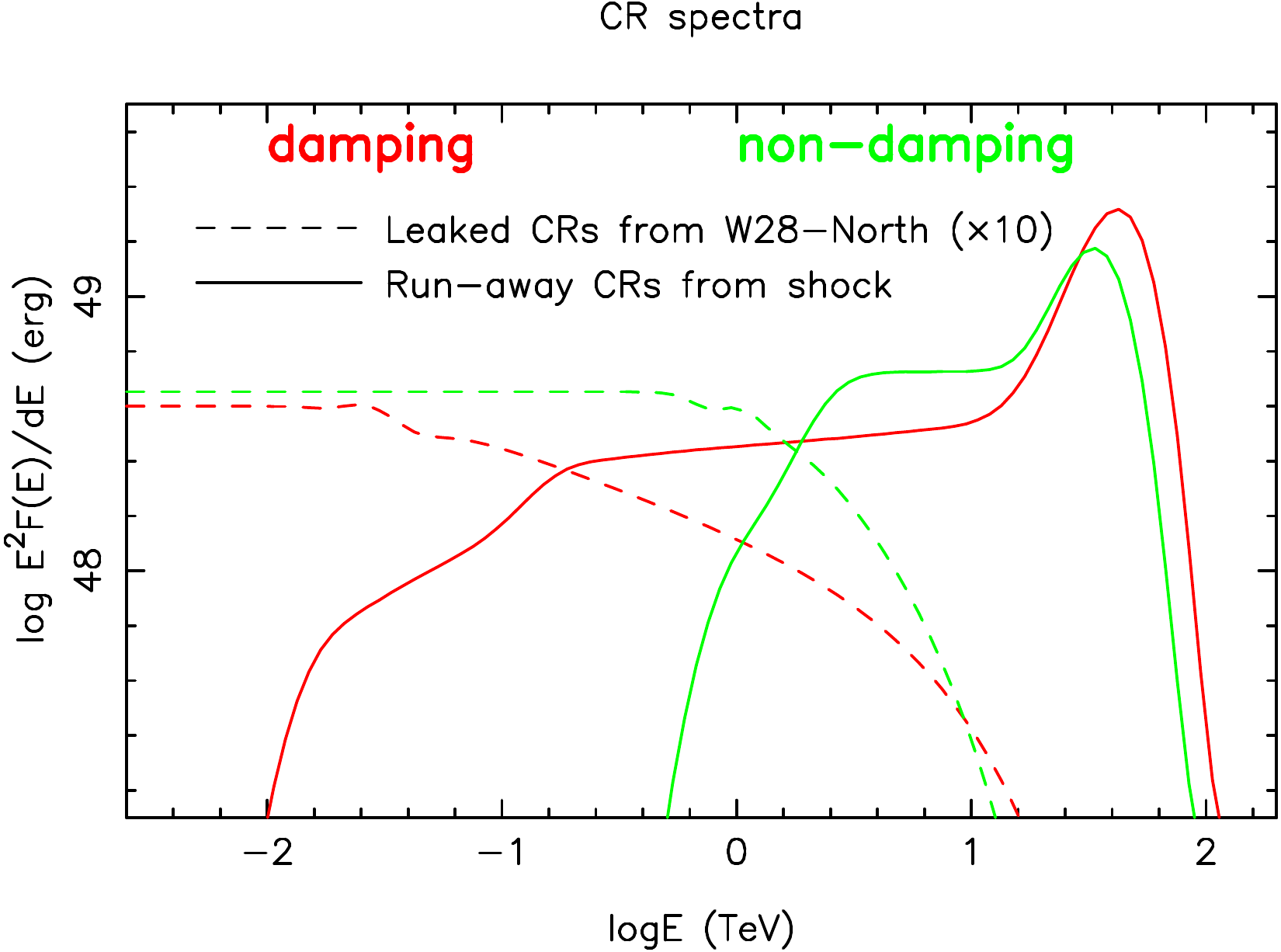}  
   \end{tabular}
   \vspace{-0.2cm}
   \caption{CR spectra of the two models. The spectrum of run-away CRs integrated till present time (37\,kyr), as well as the total leaked CRs from W28-North at 25\,kyr are marked in solid and dashed lines, respectively. The leaked CRs are made of two parts, the power-law ($\Gamma_\mathrm{CR}=-2.0$) up to $E_\mathrm{max}$ and the high energy tail. The normalization for the dashed lines has been multiplied by 10 times for graphical purposes. }
   \label{fig:CR}
\end{figure}

Knowing the run-away CR spectrum $J_\mathrm{out}$ at any given time,  we can integrate the entire SNR history as well as the whole surface of the SNR, and eventually we are able to derive the run-away CR density analytically at any location with a given diffusion coefficient, see details in section 2.4.1 of \cite{Cui2016}. Through lowering the escape energy down to 10\,GeV, which is presented in our damping model (see Fig.~\ref{fig:CR} and text below), significant amount of GeV CRs can also be released via this run-away process during the late stage of the SNR evolution. 

The damping of the magnetic waves in the upstream of the SNR, which is due to the presence of neutral atoms, becomes important in mid-aged/old SNRs \citep{Zirakashvili2017}. Hence, besides the shock-MC encounter part of the SNR -- W28-North, the other parts of this mid-aged SNR is also likely to release GeV CRs. Both the damping model and the non-damping model are explored in our work. In the non-damping model, the escape energy follows the analytical solution of the nonresonant CR streaming theory mentioned above. In the damping model, that same analytical solution only works during the early stage of SNR, when the shock velocity drops to $\sim1000\,\mathrm{km\,s^{-1}}$ ($R_\mathrm{SNR}=$4\,pc), an estimation of the escape energy in partially ionized medium of $E_\mathrm{max}=v_\mathrm{SNR}^3n_\mathrm{H}^{0.5}n_\mathrm{n}^{-1}$ by \cite{Drury1996} is adopted and is in accordance with the simulation work by \cite{Zirakashvili2017}. This number $\sim1000\,\mathrm{km\,s^{-1}}$ also gives a smooth transition from the analytical solution by \cite{Zirakashvili2008} to the estimation by \cite{Drury1996}. Here $n_\mathrm{n}$ is the number density of neutrals, which is set as $0.3\,\mathrm{cm^{-3}}$ in our damping model. 

For mid-aged/old SNRs, the CR acceleration efficiency $\eta$ is expected to decrease at low mach number shock ($M_\mathrm{s}\lesssim50$), see e.g. the hydrodynamic estimation of acceleration efficiency by \cite{Voelk1984}. Noticeably, the acceleration efficiency $\eta$ used in this work is the ratio between the energy flux of run-away CRs and the kinetic energy flux of incoming gas onto the shock. In our work, both the sound speed and Alv\'en speed in ICM is set as $15\,\mathrm{km\,s^{-1}}$ \citep{Chevalier2005}. Here we adopt a relationship of $\eta=\eta_0\exp(-(M_0/M_\mathrm{s})^{\Gamma_\mathrm{\eta}})$, where $\eta_0$ is the acceleration efficiency when the shock velocity is high, and it is set as $0.04$ and $0.03$ for the damping model and non-damping model, respectively. By choosing $\Gamma_\mathrm{\eta}=1.2-1.4$ and $M_0=4-5$ in our models, this relationship is roughly consistent with the simulation results of $0^\circ<\theta<45^\circ$ by \cite{Caprioli2014}, where $\theta$ is the angle between the normal of the shock and the magnetic field, and when $\theta\gtrsim45^\circ$ the acceleration efficiency drops significantly. These acceleration parameters adopted in our model are also constrained by the limitation of the total CR energy, see more information in section~\ref{sec:energy}.

\subsubsection{When shock encounters with MC clumps at W28-North}
\label{sec:Leaked}
The shock-MC encounter time is set at 25\,kyr ($R_\mathrm{SNR}\sim11.5\,\mathrm{pc}$) after the SN. Following the encounter, part ($\chi$) of the SNR shell located at W28-North is rapidly stalled by the dense MC clump, and only very little gas are shocked (the estimated X-ray emitting mass at W28-North is only $\sim1\,\mathrm{M_\odot}$ by \cite{Zhou2014}) before the shock is fully stalled. In fact, the shocked gas are mostly from the transition region between the diffuse gas and the dense clump. After the shock-MC encounter, a significantly slowed shock will keep propagating inside the MC clump but unable to shock the dense gas there \citep{Gabici2016}. Owing to the high density of the neutral gas, both the magnetic turbulence in the upstream and the downstream will be significantly damped, leaving the CRs confined at\&behind this part of the shock with a total number $\sim\chi N$ ready to be released, where $N$ is the total number of CRs trapped inside the SNR. During the SNR evolution history, CRs are continually carried by downstream into the interior of the SNR, with a flux $J_\mathrm{in}\sim n_0v_\mathrm{SNR}/4$, some of these trapped CRs inside SNR may re-enter the acceleration region and even escape from upstream. Noticeably, the CRs ($E\lesssim E_\mathrm{max}$) confined inside the acceleration region are almost instantly released after the shock-MC encounter, see e.g. \cite{Ohira2011}; while the GeV CRs filling in the inner region of the SNR need to be carried into MC-N by gas flow/turbulence. Assuming a magnetic field of $>10\mu G$ inside the SNR, which is consistent with the simulation results by \cite{Zirakashvili2017}, the mean Bohm diffusion distance of a 1\,TeV proton is merely $\lesssim0.3\,$pc after 10\,kyr. Once the CRs cross the stalled shock and propagate into MC-N and beyond, they will enter an environment of much higher diffusion coefficient, which is 10\% of the Galactic diffusion coefficient in our model.

Simulation work by \cite{Zirakashvili2012,Zirakashvili2017} has shown that spatial distribution of the CRs inside the SNR is dependent on the SNR age and CR diffusion coefficient (CR energy). The mid-aged/old SNRs tend to show a more homogeneous CR distribution inside SNR, while the CRs inside the young SNRs are mostly confined right behind the shock. The GeV CRs are more likely to "feel" the gas compression and more concentrated right behind the shock, while super-TeV CRs are more homogeneously filled inside the SNR.  See also the radius profiles of CR pressure and gas density inside different SNRs by \cite{Zirakashvili2012}. 

Additional to those CRs in the acceleration region who are immediately released into MC-N after the shock-MC encounter, the GeV CRs located in the inner region of the SNR, who "feel" the gas compression, can also be carried into MC-N by the downstream flows/turbulence. 
In our model, the CR releasing from W28-North (including both the CRs from acceleration region and the CRs carried by flow) is arbitrarily chosen as an instantaneous process. Hence, this releasing time (12\,kyr ago) actually functions like an averaged releasing time, and the total energy of leaked CRs from W28-North is normalized later in our modelling.

Similar to the spectrum of the CRs confined inside the SNR in the simulation work by \cite{Zirakashvili2017}, the one in our model (also the spectrum of leaked CRs), as seen in Fig.~\ref{fig:CR}, is arbitrarily made of two components: First, a low energy part with a power-law index of $\Gamma_\mathrm{CR}=-2.0$ and a cutoff energy of $E_\mathrm{max}$, which represent the CR spectrum in/near the acceleration region; Second, a high energy tail extending beyond $E_\mathrm{max}$ with a power-law index of $\Gamma_\mathrm{CR}=-2.3(-2.5)$ and a cutoff energy of $10(5)\,$TeV for the damping(non-damping) model. In our models, $\chi$ represents how much percentage of the SNR shell is stalled by the MC-N at 25\,kyr. After the MC-shock encounter, part ($\chi$) of the shock start to release all its GeV CRs, while the other $1-\chi$ GeV CRs will mostly remain behind their local shock. $\chi$ is a free parameter here, however, it is constrained by the limitation of the total CR energy. E.g., by choosing $\chi=10\%$ we would derive a total CR energy in the SNR of $\sim4\%/11\%\,\mathcal{E}_{51}$ 12\,kyr ago in our damping/non-damping model. 

At 25\,kyr after the SN, the energy of TeV CRs trapped inside the SNR only account for $\lesssim7\%$ of the energy of total trapped CRs in our models. Unlike the GeV CRs, all of the diffuse super-TeV CRs trapped in the entire SNR (rather than 10\% of the SNR) are able to be gradually released through the leaking hole of W28-North. In our models, the run-away CRs from the early SNR stages dominate the higher energy $\gamma$-ray emission at MC clumps. Therefore, for simplicity, part ($\chi$) of the TeV CRs is bind together with the GeV CRs and released instantaneously from W28-North.

\subsubsection{Sea of the Galactic CR}

The spatial distribution of Galactic sea CRs, especially the GeV sea CRs, could be quite inhomogeneous. Naturally, the GeV CRs are likely to be concentrated near the star forming regions, e.g. the spiral arms. Through studying the Fermi-LAT data and the gas density in the entire Galaxy, \cite{Acero2016} has derived a rough radius profile of CR density/spectrum in the Galactic plane. Considering that SNR\,W28 lies inside the Galactic plane with a distance to the Galactic center of $\sim5\,$kpc, using this radius profile one can obtain the spectrum of its local CR sea (CR spectrum index $\Gamma_\mathrm{CR,sea} \sim$ -2.55 to -2.72, energy density $U_\mathrm{CR,sea}\sim1.1\,\mathrm{eV\,cm^{-3}}$), which is similar to the one detected on Earth. However, in our models, we explore scenarios with CR sea densities lower than the averaged local one mentioned above, in order to better fit the $\lesssim1\,\mathrm{GeV}$ observations (especially for 240A).

 \subsection{Results and discussions} \label{sec:result}
\begin{figure}[!h]
   \centering
\includegraphics[width=16cm]{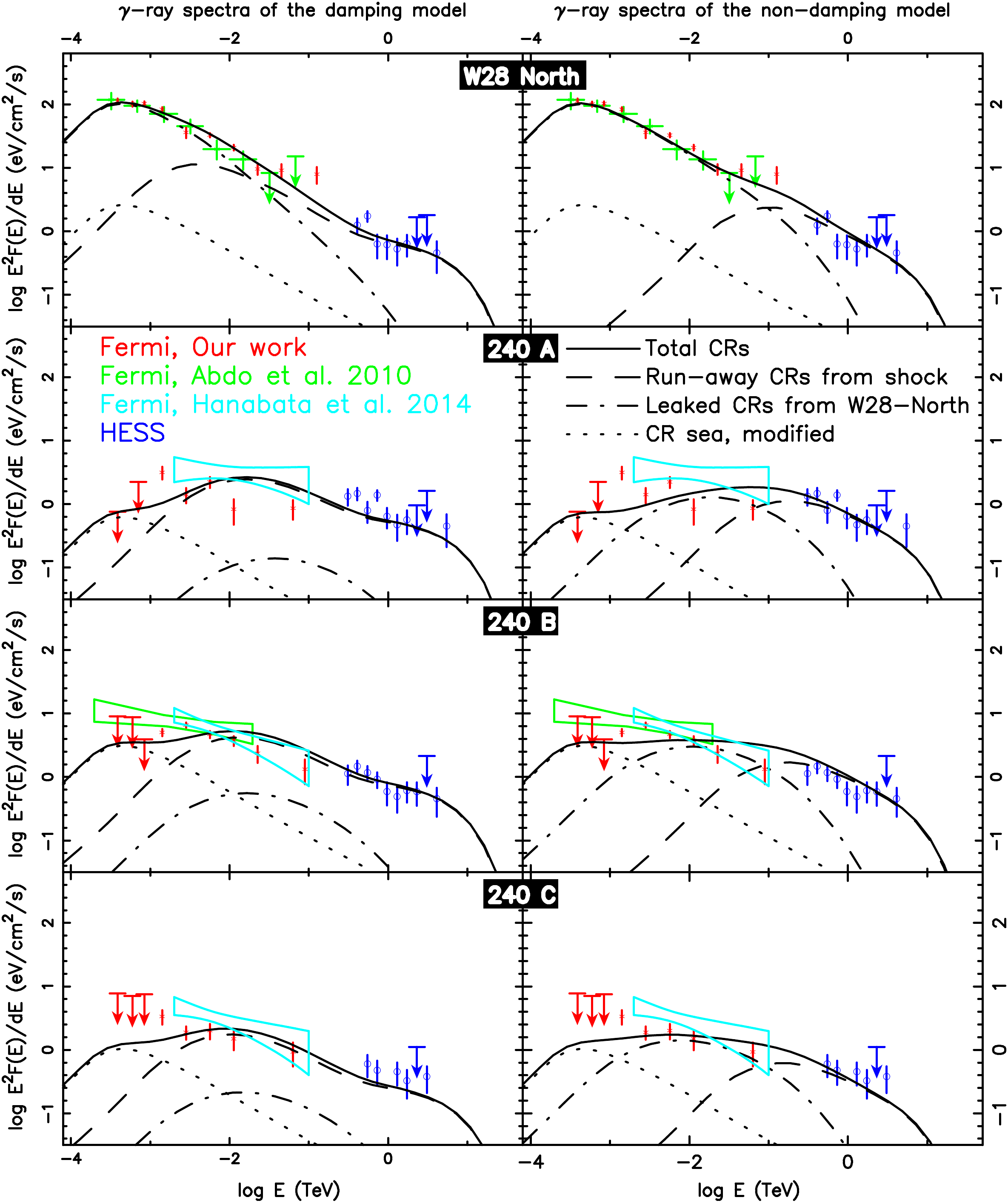} 
   \vspace{-0.2cm}
   \caption{Broad band fit to the $\gamma$-ray emission from the sources HESS\,J1801-233 (W28 north), HESS\,J1800-240\,A, B, and C (top to bottom). The left/right panels are the fitting results of the damping/non-damping model, which are based on our own Fermi-LAT data points (red stars) and the H.E.S.S. data (blue circles). The observational data are marked in the same way as the ones in Fig.~\ref{fig:pre_model}. In each panel, the hadronic $\gamma$-ray produced via the run-away CRs from the shock, the released CRs from W28-North, the sea of Galactic CRs, and the total CRs are shown in dashed, dash-dotted, dotted, and solid lines, respectively. The density of CR sea at 240\,A (W28-North, 240\,B\,\&\,C) is modified to 14\% (50\%) of the one detected on Earth.  }
   \label{fig:spec2}
\end{figure}

In this work, we try to reproduce the GeV-TeV spectrum at MC-N, MC-A, MC-B and MC-C simultaneously in one model. To explain the GeV observation, the key strategy in our model is realized by introducing released GeV CRs from the broken shell at W28-North rather than decreasing the energy of run-away CRs down to 1\,GeV.
As seen in Fig.~\ref{fig:spec2}, the GeV-TeV observations in/around W28 are explained except for 240\,A. In our models, the TeV emission in/around W28 is dominated by run-away super-TeV CRs released during the early stages of the SNR, while the leaked CRs from W28-North dominate almost the entire GeV band of W28-North and also contribute to the $\sim3-100$ GeV emission at the distant MC clumps. The $\lesssim$3\,GeV emission at the distant MC clumps are explained by the local CR sea.
For 240\,B, the reproduced GeV flux at around 2\,GeV is slightly lower than the observational data, and this can be due to the contamination from the south blob. For 240\,C, besides reducing the south blob, we could also use a higher density of CR sea to better fit the $\lesssim$1\,GeV emission (using the CR sea density detected on Earth, which is not shown in this paper).
For 240\,A, our reproduced GeV spectrum can not fit the sharp peak of 2\,GeV shown in the observational data, also 240\,A is spatially far away from this south blob. 
More discussions about our model and results are listed in following subsections.

\subsubsection{The diffusion coefficient and diffusion distances}
\label{sec:diffusion}
In the reproduction of the $\gamma$-ray emission at MC-N, MC-A,\,B,\,C, the diffusion coefficient for the entire space outside the SNR is fixed as $D(E)=10^{27}(E/10\,\rm{GeV})^\delta cm^2/s,\ \delta=0.5$, mostly based on the requirement of the fitting of the GeV data of W28-North and the TeV data of all the $\gamma$-ray sources. Thus the fine tuning of the distances between these MC clumps and the CR sources is performed for a better fitting result, see table~\ref{table:Distance}, the 3-dimensional distances between the MC clumps and the CR sources in our model satisfy the triangle relationship among the three objects: W28-North, SNR center, and each MC clump. 
 
For GeV CRs, their spatial distribution are sensitive to the distances to the CR sources. The distances of MC-A, B, C to W28-North are around 25-35\,pc in our model, and 25\,pc is around the projected distance between W28-North and MC-B,C. Our model face difficulties to efficiently bring $\lesssim$10\,GeV CRs from the SNR to MC-A, MC-B, who has shown $\gamma$-ray spectra peaks at 1.4\,GeV and 2.8\,GeV, respectively. Adding the CR sea could help moving the peaks of reproduced $\gamma$-ray emission to lower energy, see the dotted lines and solid lines in Fig.~\ref{fig:spec2}. 

In case of 240\,A, if we put MC-A much closer to the SNR, e.g. 20\,pc (same effect as using a higher diffusion coefficient in the entire space), although we could move the peak to lower energy, but the reproduced 10-100\,GeV flux will be too high to explain the observation as well. This over production of 10-100\,GeV flux also happens in cases of MC-B,C. By modifying the diffusion coefficient, one solution for such situation is to introduce a diffusion coefficient with a smaller power-law index (compared to 0.5 for a Galactic one), which allow the 1-100\,GeV CRs to diffuse relative faster. 
Another solution would be introducing a diffusion coefficient spectrum wth broken power-laws: the power-law index of the diffusion coefficient of 1-100\,GeV CRs can be smaller than 0.5. Interestingly, the derived diffusion coefficients inside a Kolmogorov's turbulence do show such kind of broken power-laws, see e.g. the numerical simulation results by \cite{Casse2002,Fatuzzo2010}. However, a power-law index of the diffusion coefficient smaller than 0.5 would require a softer spectrum of the leaked GeV CRs ($\Gamma_\mathrm{CR}<-2.0$), to explain the GeV spectrum of W28-North. Furthermore, an anisotropic or/and inhomogeneous diffusion environment could also help our modelling, e.g. fast diffusing tubes based on the large magnetic structure. Overall, these alternative diffusion scenarios which may need a total modification of our model, including the acceleration and releasing of the GeV-TeV CRs, will be left to future work.

\subsubsection{An inhomogeneous CR sea near W28?}
Due to the long distances from W28-North to MC-A, B, C ($\gtrsim20$\,pc), $1-100$\,GeV CRs leaked from W28-North are not able to efficiently reach these MC clumps at present time with the diffusion coefficient used in our model. Hence, the CR sea is expected to dominate the $\lesssim10$\,GeV $\gamma$-ray emission at these three MC clumps. The density of CR sea (5\,kpc from Galactic center) observed by \cite{Acero2016} contradicts our Fermi-LAT discovery -- the non-detection of $\lesssim1\,$GeV $\gamma$-ray at MC-A. These observed MC clumps cover a distance to Earth ranging from 2\,kpc to 4\,kpc \cite{Ah2008}, however, putting them further from Earth (e.g. 4\,kpc instead of 2\,kpc) (power sources other than W28 could be introduced as well to explain the higher energy band) only leads to a situation that the MC clumps are put closer to the Galactic center ($3-4\,$kpc), where a $\sim3$ times higher density of CR sea than the one detected on Earth is suggested by \cite{Acero2016}. Noticeably, the diffuse background in the Fermi analysis is very important at this low energy band, and we can not rule out that our non-detection of $\lesssim1\,$GeV $\gamma$-ray at MC-A could be the result of an excessive background reduction, see our background in Fig.~\ref{fig:background}.

In summary, if the low $<$1GeV flux at 240\,A is true, an inhomogeneous distribution of the $<10\,$GeV CR sea becomes necessary. The density of CR sea adopted in Fig.\ref{fig:spec2} is half of the averaged local one, except for 240\,A who adopt a much lower one of 14\% of the averaged local one.

\subsubsection{The run-away GeV CRs}
\label{sec:1GeV}
As described in the introduction section, considering the ionization evidences found at MC-N, we believe the CRs with energy down to $<1\,$GeV have already leaked through the broken shell at W28-North and dominate the GeV emission there. By lower the escape energy, run-away CRs can also contribute significantly to the GeV emission at W28-North. Nonetheless, in the damping theory, the ionization ratio in the ICM which functions as a key factor on the escape energy is unknown. With an ICM density of $\lesssim10\mathrm{cm^{-3}}$, an ICM temperature of $T\gtrsim10^4\,$K, which is just above the ionization temperature of H, is needed to confine the dense clumps and to support the cloud against gravitational collapse \citep{Blitz1993, Chevalier1999}.  Therefore, in our damping model, the neutral density of the upstream ICM is set as low as $0.3\,\mathrm{cm^{-3}}$ and the final escape energy of the $100\,\mathrm{km\,s^{-3}}$ shock at present time is 10\,GeV. To further decrease the escape energy to 1\,GeV and allow the run-away CRs (from the shocks other than W28-North) to dominate the $<1\,$GeV $\gamma$-ray emission at MC-N, one requires a neutral density $\sim3\,\mathrm{cm^{-3}}$ in the ICM following the estimation by \cite{Drury1996}. This requirement is slightly contrary to the observed H{\sc i} density upper-limits of $1.5-2\,\mathrm{cm^{-3}}$ by \cite{Velazquez2002}.

\subsubsection{Progenitors other than 8\,$\mathrm{M_\odot}$}
In our SNR evolution model of W28, we have only explored a core collapse SN scenario which delivers a homogeneous pre-SN environment. This scenario has successfully explained the observed shock velocity at present time and the measured gas density near the SNR. Here we discuss another three hypothesis which has more massive progenitors and more complex pre-SN wind environments. 
\begin{enumerate}
\item {\it A big pre-SN wind bubble, radius $R_\mathrm{b}>13\,\mathrm{pc}$}. Due to the low shock velocity at present and the lack of non-thermal X-ray, SNR\,W28 is unlikely still evolving inside the pre-SN wind bubble who has a typical density of $\sim0.01\,\mathrm{cm^{-3}}$. 
\item {\it A medium pre-SN wind bubble, radius $13\,\mathrm{pc}>R_\mathrm{b}\gtrsim10\,\mathrm{pc}$}. If SNR\,W28 encounters the pre-SN bubble shell relative recently, we are more likely to see a shell structure of thermal X-ray rather than the observed features: A diffused one in the center and a bright one at W28-North.
\item {\it A small pre-SN wind bubble, radius $R_\mathrm{b}\lesssim10\,\mathrm{pc}$}. If SNR\,W28 has already swept the pre-SN bubble shell and is expanding in the ICM before it enters the radiation-dominated stage, similar shock velocity ($v_\mathrm{SNR}\sim100\,\mathrm{km\,s^{-1}}$) at present time is expected (adopting the same SN energy $\mathcal{E}_\mathrm{ej}=\mathcal{E}_{51}$). This is due to that the final shock velocity during the Sedov stage is mostly sensitive to the total swept mass but not to the radius distribution of the gas, see e.g. the scenario $8\mathrm{M_\odot}$ and scenario $15\mathrm{M_\odot}$ in \cite{Cui2016}. In this new SNR scenario which is not explored in this work, the GeV CR leaking process which happens after the sweeping of the pre-SN bubble shell is expected to be similar to that in the our scenario, however, the releasing of super-TeV CRs which happens mostly in the early stage of the SNR evolution will be different. 
\end{enumerate}

\subsubsection{The total energy of accelerated CRs at SNR W28}
\label{sec:energy}
Since our SNR evolution history is fixed, then the key normalization factor on the total accelerated CR energy becomes the acceleration efficiency, see table~\ref{table:Models}.
Adapting the $\eta_\mathrm{esc}=0.04/0.03$ in our damping/non-damping model, at the shock-MC encounter time ($25$\,kyr), the SNR has already released most of its run-away CRs with a total energy about $9.1\%/6.6\% \,\mathcal{E}_{51}$. After parts ($\chi$) of the SNR have encountered with MC-N (25\,kyr), about $0.4\%/1.1\% \,\mathcal{E}_{51}$ CRs are released from the W28-North in the damping/non-damping model. If we follow the concept by \cite{Zirakashvili2012,Zirakashvili2017} that most of the accelerated CRs have escaped the SNR through run-away CRs for a mid-aged/old SNR, then our total CR energy inside SNR at 25kpc should be smaller than the one of run-away CRs, which leads to a $\chi\gtrsim4.3\%/16.7\%$ in the damping/non-damping model. After the shock-MC collision, only the other $1-\chi$ part of the shell are still releasing run-away CRs, $\chi$ is set as 10\%/15\% in our damping/non-damping model.

\subsubsection{The CR leaking process at W28-North}
Based on the X-ray ionization time at W28-North by \cite{Zhou2014}, we simply adopt an instantaneously releasing time (12\,kyr ago from present) to set free the CRs at W28-North. In fact, as described in section~\ref{sec:Leaked}, only CRs from the acceleration region are released instantaneously, while the GeV/TeV CRs from the inner region are gradually injected into MC-N advectively/diffusively. After the shock-MC encounter, the tip of MC-N which is buried inside the SNR will keep being blown by a downstream wind (flow) with a velocity of $v_\mathrm{flow}$, where $v_\mathrm{flow}$ is a little smaller than the current shock velocity (at 25\,kyr, $v_\mathrm{flow}\lesssim150\,km\,s^{-1}$). However, this advection-dominated injection of GeV CRs by the downstream flow will decrease with time, because: Firstly, the CR density is gradually decreasing from the region right behind the shock to the inner region of the SNR, see e.g. the simulation result of mid-aged/old SNRs by \cite{Zirakashvili2012}, whose CR density is halved at 10\%\,$R_\mathrm{SNR}$ downstream from the shock front; Secondly, the flows that are falling onto MC-N will decrease when the gas pressure behind the stalled shock reaches new balance and the eddies stirred by MC-N decay in several crossing times $t_\mathrm{c}=L_\mathrm{ed}/v_\mathrm{flow}$, where $L_\mathrm{ed}$ is the length of the eddies and should be comparable to the size of the MC-N tip which is embedded in the SNR ($t_\mathrm{c}\sim10^4\,$year when $L_\mathrm{ed}\sim1$\,pc).
To fully unveil the leaking process at W28-North, a MHD simulation is required in future work, to obtain the flow/turbulence details as well as the diffusion environments in/around this shock-MC encounter region. 

Noticeably, this shock-MC encounter at W28-North could be a more complex process than our simple ``colliding onto a flat wall" model, considering that the 3-dimensional structure of this MC-N could be very complex. This complex encounter process could last for quite a long period, e.g., with an expanding velocity $\sim100\,\mathrm{km\,s^{-1}}$, SNR W28 would need $\gtrsim10^4$\,year to fully swallow MC-N, whose size is $\gtrsim1\,$pc.

\section{Summary} \label{sec:summary}
This work is motivated by the exciting discovery of the GeV-TeV emission in/around SNR W28 several years ago, which seems to indicate a picture that the super-TeV CRs released at early stage are evenly spread into the nearby environment and explain well the spatial coincidence between the TeV emission and the molecular clouds, meanwhile, the GeV CRs released from W28-North (or from the entire SNR) are more concentrated near the CR releasing source, and this eventually explains why MC-N is the most illuminated source in Fermi-LAT skymap with a peak energy down to $\lesssim$0.3\,GeV. To further testify this picture, Firstly, we re-analysis the Fermi-LAT data of W28 and extend the GeV spectra at MC-A, B, C down to 0.3\,GeV; Secondly, we explore a hadronic model involving leaked GeV CRs from a broken shell. Detailed summary of these two work are described in following text.
\begin{itemize}
\item {\it The Fermi-LAT analysis of W28.} Using the 9 years Fermi-LAT data and the most updated Fermi Science Tools -- v10r0p5 package, we re-explored the GeV counterparts of HESS\,J1801-233, 240\,A,\,B\,\&\,C, as well as newly discovered GeV source at west of W28 -- 3FGL J1758.8-2346 (Source-W). In morphological study, our 10-200\,GeV skymap is in accordance with the H.E.S.S. skymap, and the position of Source-W is slightly revised when compared to the one by the previous work. In spectral study, the spectra of the southern GeV counterparts has shown discontinuities at $\sim1\,$GeV -- a sudden flux increase at the $\lesssim1\,$GeV band of 240\,B\,\&\,C, while a non-detection at the $\lesssim1\,$GeV band of 240\,A. Further morphological study triggered by these discontinuities has discovered a 0.5-1\,GeV blob (south blob) located to the south of 240\,B\,\&\,C. This south blob could be due to either an unknown GeV source or an unclean background reduction. Ultimately, only upper-limits of $<1\,$GeV band at 240\,B\,\&\,C are drawn in our final results.
\item {\it The hadronic model involving leaked CRs from the broken shell.} The main goal of our modelling work is to testify that whether the $\lesssim10$\,GeV emission in W28 can be explained by the leaked CRs from the shock-MC collision -- W28-North, rather than by run-away CRs with energy down to 1\,GeV. Following the multi-wavelength observational constrains, a core collapse SN scenario with a progenitor mass of $8\,M_\mathrm{\odot}$ is adopted in our model, which derives a shock velocity of $\sim100\,\mathrm{km\,s^{-1}}$ at present time (37\,kyr). Considering the mid-age of SNR\,W28, both the damping of the magnetic waves caused by neutral atoms and the decreasing acceleration efficiency caused by low shock speed are taken into account in our model.
The GeV-TeV emissions in/around W28, except for 240\,A, are explained in our hadronic model. Our model consists of three CR sources: the run-away CRs escaped from the upstream of strong shocks, the leaked CRs from the broken shell -- W28-North, and the Galactic CR sea. The releasing of run-away CRs follows the nonresonant acceleration theory developed by \cite{Zirakashvili2008}, which can provide the escape energy during entire the SNR history. The escape energy of run-away CRs ranges from $\sim50$\,TeV to 10\,GeV/1\,TeV in our damping/non-damping model. We assume that the SNR encounters with MC clumps at W28-North $1.2\,$kyr ago, and following this encounter, the leaked CRs (mostly are GeV CRs with energy below the escape energy) are released instantaneously. Due to the long distances of 240\,A,\,B,\,C to the SNR, their $\lesssim10$\,GeV emission are dominated by the local CR sea in our model. The finding of the non-detection of $<1\,$GeV emission at 240\,A is sensitive to the diffuse background, if this finding is true, an inhomogeneous density of CR sea (even in the case MC-A is 4\,kpc from Earth) is expected, which is necessary to provide a much lower CR sea density than the averaged one at a distance of 3-5\,kpc to the Galactic center. 
\end{itemize}

\section*{Acknowledgement}
We like to thank Vladimir Zirakashvili for the helpful advises on the cosmic ray acceleration theory. This work is supported by the National Science Foundation of China (NSFC) grants 11633007, 11661161010, and U1731136.




\clearpage

\clearpage

\clearpage

\clearpage

\clearpage

\clearpage

\begin{table}
	\caption{The values of $-ln(likelihood)$ in 10-200 GeV, where the radius of the uniform-disk source 3FGL J1801.3-2326e is changed to different values.}
	\begin{center}
				\begin{tabular}{lc}
			\hline\hline
			Radius of extension (deg) & $-ln(likelihood)$ \\ \hline
			0.1                       & 85302.65878385  \\
			0.2                       & 85281.56893407  \\
			0.3                       & 85250.41629475  \\
			0.33                      & 85247.72361     \\
			0.34                      & 85247.39720513  \\
			0.345                     & 85247.39159966  \\
			0.35                      & 85247.3950346   \\
			0.36                      & 85247.81674768  \\
			0.4                       & 85249.53678938 \\ \hline
		\end{tabular}
	\end{center}
	\label{Ext}
\end{table}

\clearpage

\begin{table}
	\caption{0.3-250 GeV spectral properties.}
	\begin{center}
		\footnotesize{\begin{tabular}{lccccc}
			\hline\hline
			& \emph{Fermi} J1801.4-2326     &       240\,A  &       240\,B   &       240\,C   &       Source-W       \\ \hline
			\multicolumn{6}{c}{PL}            \\ \hline
			$\Gamma$       & 2.422   $\pm$ 0.009 & 2.164 $\pm$ 0.061 & 2.209  $\pm$ 0.036 & 2.455  $\pm$ 0.070 & 2.133    $\pm$ 0.059 \\
			Flux ($10^{-9}$\,cm$^{-2}$\,s$^{-1}$) & 315.816 $\pm$ 2.965 & 8.282 $\pm$ 1.066 & 21.188 $\pm$ 1.267 & 15.920 $\pm$ 1.379 & 8.456    $\pm$ 0.998 \\
			TS          & 19007.8 &       164.7 &       642.1  &       236.5  &       168.0          \\ \hline
			\multicolumn{6}{c}{BKPL}          \\ \hline
			$\Gamma_1$      & 2.095   $\pm$ 0.025 & 0.530 $\pm$ 0.721 & 2.245  $\pm$ 0.145 & 2.984  $\pm$ 0.182 & -0.086   $\pm$ 0.849 \\
			$\Gamma_2$      & 2.629   $\pm$ 0.019 & 2.371 $\pm$ 0.092 & 2.200  $\pm$ 0.049 & 2.237  $\pm$ 0.088 & 2.416    $\pm$ 0.098 \\
			$E_\mathrm{b}$ (MeV)    & 1000    &       1000  &       1000   &       1000   &       1000           \\
			Flux ($10^{-9}$\,cm$^{-2}$\,s$^{-1}$) & 308.367 $\pm$ 3.021 & 5.912 $\pm$ 1.232 & 21.346 $\pm$ 1.417 & 17.081 $\pm$ 1.375 & 5.640    $\pm$ 1.083 \\
			TS          & 19216.0 &       178.3 &       642.1  &       245.8  &       190.2    \\ \hline      
		\end{tabular}}
	\end{center}
	\label{spectral1}
\end{table}

\clearpage

\begin{table}
	\caption{Spectral properties  in narrowed energy bands.}
	\begin{center}
		\footnotesize{\begin{tabular}{lccc}
			\hline\hline
			& \emph{Fermi} J1801.4-2326  &       240\,B                       & 240\,C       \\ \hline
			Energy band (GeV) & -- & 0.3-0.75 &       0.3-1       \\ \hline
			\multicolumn{4}{c}{\underline{PL}}       \\
			$\Gamma$      & --       & 2.154    $\pm$ 0.386 & 2.113 $\pm$ 0.212 \\
			Flux ($10^{-9}$ ph cm$^{-2}$ s$^{-1}$)   & --    & 18.51    $\pm$ 1.40  & 18.81 $\pm$ 1.37  \\
			TS       & --         & 211.4    &       244.7       \\ \hline
			Energy band (GeV) &       1-250& 0.75-250 &       1-250       \\ \hline
			\multicolumn{4}{c}{\underline{PL}}       \\
			$\Gamma$      & 2.637 $\pm$ 0.021       & 2.238    $\pm$ 0.045 & 2.237 $\pm$ 0.101 \\
			Flux ($10^{-9}$ ph cm$^{-2}$ s$^{-1}$)    & 61.092 $\pm$ 0.881   & 7.540    $\pm$ 0.415 & 2.263 $\pm$ 0.276 \\
			TS     &     8413.9      & 572.5    &       111.0       \\
			\multicolumn{4}{c}{\underline{BKPL}}       \\
			$\Gamma_1$      & 2.650 $\pm$ 0.022      & 1.769    $\pm$ 0.149 & 2.159 $\pm$ 0.203 \\
			$\Gamma_2$      & 2.123 $\pm$ 0.216      & 2.469    $\pm$ 0.088 & 2.371 $\pm$ 0.314 \\
			$E_\mathrm{b}$ (MeV)    & 33262 $\pm$ 2354      & 2780     $\pm$ 837   & 8686  $\pm$ 4896  \\
			Flux ($10^{-9}$ ph cm$^{-2}$ s$^{-1}$)   & 61.138 $\pm$ 0.881    & 6.978    $\pm$ 0.455 & 2.212 $\pm$ 0.408 \\
			TS    &      8418.0      & 585.8    &       111.3 \\ \hline      
		\end{tabular}}
	\end{center}
	\label{spectral2}
\end{table}

\clearpage

\clearpage

\clearpage

\begin{table}
	\caption{The distances (pc) between MCs and CR sources}
	\begin{center}
		\footnotesize{\begin{tabular}{lcccc}
			\hline\hline
			& MC-N ($5\,\mathrm{M_{4}}$\tablenote[1]{ Here $\mathrm{M_{4}} = 10^{4} \mathrm{M_\odot}$ })    &       MC-A ($4.3\,\mathrm{M_4}$) &       MC-B ($6\,\mathrm{M_4}$)   &       MC-C ($2\,\mathrm{M_4}$)      \\ \hline
			\multicolumn{5}{c}{Damping}            \\ \hline
			SNR  center    & 13   & 35   & 31 & 27  \\
			W28-North & 0$\sim$1  & 37 & 29 & 28  \\ \hline
			\multicolumn{5}{c}{Non-damping}          \\ \hline
		SNR  center    & 13   & 35   & 28 & 27  \\
			W28-North & 0$\sim$1  & 33 & 26 & 25  \\ \hline
  \hline      
		\end{tabular}}
	\end{center}
	\label{table:Distance}
\end{table}

\clearpage
\begin{table*}
\caption{Parameters in the two models  
}  
\begin{center}          
\begin{tabular}{ | c | c | c | c | c | c | c | c |}    
\hline       
Models \tablenote[1]{ Both two models share the same SNR evolution history, diffusion environment.}  & $\chi$ & $\eta_\mathrm{0} $ & $M_0$ & $\Gamma_\mathrm{\eta}$ &$E_\mathrm{max}$\tablenote[2]{ The escape energy at present time (37\,kyr).}  (TeV) &$E_\mathrm{CR, run}$\tablenote[3]{ The total energy of run-away CRs at present time (37\,kyr).} ($\mathcal{E}_{51}$) & $E_\mathrm{CR, leak}$\tablenote[4]{ The total energy of CRs leaked through W28-North at the shock-MC encounter time (25\,kyr). } ($\mathcal{E}_{51}$) \\ 
\hline 
Damping & 10\% &0.04& 4 & 1.4 & 0.012  & 9.2\% & 0.4 \%\\ \hline
Non-damping & 15\% &0.03& 5 & 1.2 &0.72 & 6.7\% &  1.1\% \\
\hline                 
\end{tabular}
\end{center}
\label{table:Models} 
\end{table*}

\clearpage

\listofchanges


\begin{thebibliography}{}

\bibitem[Abdo et al.(2010)]{Abdo2010} Abdo, A.~A., Ackermann, 
M., Ajello, M., et al.\ 2010, \apj, 718, 348



\bibitem[Acero et al.(2015)]{Acero2015a} Acero, F., Ackermann, M., Ajello, M., et al.\ 2015, \apjs, 218, 23

\bibitem[Acero et al.(2016)]{Acero2016} Acero, F., Ackermann, M., Ajello, M., et al.\ 2016, \apjs, 223, 26


\bibitem[Aharonian et al.(2008)]{Ah2008} Aharonian, F., Akhperjanian, A.~G., Bazer-Bachi, A.~R., et al.\ 2008, \aap, 481, 401 


\bibitem[Bell(2004)]{Bell2004} Bell, A.~R.\ 2004, \mnras, 353, 
550



\bibitem[Bisnovatyi-Kogan \& Silich(1995)]{Bisnovatyi1995} Bisnovatyi-Kogan, G.~S., \& Silich, S.~A.\ 1995, Reviews of Modern Physics, 67, 661 

\bibitem[Blitz(1993)]{Blitz1993} Blitz, L.\ 1993, Protostars and Planets III, 125

\bibitem[Bohigas et al.(1983)]{Bohigas1983} Bohigas, J., Ruiz, M.~T., Carrasco, L., Salas, L., \& Herrera, M.~A.\ 1983, Revista Mexicana de Astronomia y Astrofisica, 8, 155 


\bibitem[Brogan et al.(2006)]{Brogan2006} Brogan, C.~L., Gelfand, J.~D., Gaensler, B.~M., Kassim, N.~E., \& Lazio, T.~J.~W.\ 2006, \apjl, 639, L25 



 \bibitem[Casse et al.(2002)]{Casse2002} Casse, F., Lemoine, M., \& Pelletier, G.\ 2002, \prd, 65, 023002

\bibitem[Cioffi et al.(1988)]{Cioffi1988} Cioffi, D.~F., McKee, C.~F., \& Bertschinger, E.\ 1988, \apj, 334, 252


\bibitem[Chevalier(1982)]{Chevalier1982} Chevalier, R.~A.\ 1982, 
\apj, 259, 302

\bibitem[Chevalier(1999)]{Chevalier1999} Chevalier, R.~A.\ 1999, \apj, 511, 798

\bibitem[Chevalier(2005)]{Chevalier2005} Chevalier, R.~A.\ 2005, \apj, 619, 839 

\bibitem[Claussen et al.(1999)]{Claussen1999} Claussen, M.~J., Goss, W.~M., Frail, D.~A., \& Desai, K.\ 1999, \apj, 522, 349


\bibitem[Cui et al.(2016)]{Cui2016} Cui, Y., P{\"u}hlhofer, G., \& Santangelo, A.\ 2016, \aap, 591, A68 

\bibitem[Caprioli \& Spitkovsky(2014)]{Caprioli2014} Caprioli, D., \& Spitkovsky, A.\ 2014, \apj, 783, 91 

\bibitem[O'C Drury et al.(1996)]{Drury1996} O'C Drury, L., Duffy, P., \& Kirk, J.~G.\ 1996, \aap, 309, 1002


\bibitem[Fatuzzo et al.(2010)]{Fatuzzo2010} Fatuzzo, M., Melia, F., Todd, E., \& Adams, F.~C.\ 2010, \apj, 725, 515 

\bibitem[Frail et al.(1994)]{Frail1994} Frail, D.~A., Goss, W.~M., \& Slysh, V.~I.\ 1994, \apjl, 424, L111

\bibitem[Gabici et al.(2010)]{Gabici2010} Gabici, S., Casanova, 
S., Aharonian, F.~A., 
\& Rowell, G.\ 2010, SF2A-2010: Proceedings of the Annual meeting of the French Society of Astronomy and Astrophysics, 313 

\bibitem[Gabici \& Aharonian(2016)]{Gabici2016} Gabici, S., \& Aharonian, F.\ 2016, European Physical Journal Web of Conferences, 121, 04001 

\bibitem[Hanabata et al.(2014)]{Hanabata2014} Hanabata, Y., Katagiri, H., Hewitt, J.~W., et al.\ 2014, \apj, 786, 145

\bibitem[Hewitt \& Yusef-Zadeh(2009)]{Hewitt2009} Hewitt, J.~W., \& Yusef-Zadeh, F.\ 2009, \apjl, 694, L16


\bibitem[Inoue et al.(2012)]{Inoue2012} Inoue, T., Yamazaki, R., Inutsuka, S.-i., \& Fukui, Y.\ 2012, \apj, 744, 71


\bibitem[Li \& Chen(2010)]{Li2010} Li, H., \& Chen, Y.\ 2010, \mnras, 409, L35 

\bibitem[Long et al.(1991)]{Long1991} Long, K.~S., Blair, W.~P., Matsui, Y., \& White, R.~L.\ 1991, \apj, 373, 567 

\bibitem[Lynds \& Oneil(1985)]{Lynds1985} Lynds, B.~T., \& Oneil, E.~J., Jr.\ 1985, \apj, 294, 578

\bibitem[Marquez-Lugo \& Phillips(2010)]{Marquez2010} Marquez-Lugo, R.~A., \& Phillips, J.~P.\ 2010, \mnras, 407, 94 

\bibitem[Maxted et al.(2016)]{Maxted2016} Maxted, N.~I., de Wilt, P., Rowell, G.~P., et al.\ 2016, \mnras, 462, 532


\bibitem[Nadezhin(1985)]{Nadezhin1985} Nadezhin, D.~K.\ 1985, \apss, 112, 225 

\bibitem[Nakamura et al.(2014)]{Nakamura2014} Nakamura, R., Bamba, A., Ishida, M., et al.\ 2014, \pasj, 66, 62 

\bibitem[Neufeld et al.(2007)]{Neufeld2007} Neufeld, D.~A., Hollenbach, D.~J., Kaufman, M.~J., et al.\ 2007, \apj, 664, 890

\bibitem[Nicholas et al.(2011)]{Nicholas2011} Nicholas, B., Rowell, 
G., Burton, M.~G., et al.\ 2011, \mnras, 411, 1367 

\bibitem[Nicholas et al.(2012)]{Nicholas2012} Nicholas, B.~P., 
Rowell, G., Burton, M.~G., et al.\ 2012, \mnras, 419, 251 

\bibitem[Ohira et al.(2011)]{Ohira2011} Ohira, Y., Murase, K., \& Yamazaki, R.\ 2011, \mnras, 410, 1577

\bibitem[Ostriker 
\& McKee(1988)]{Ostriker1988} Ostriker, J.~P., \& McKee, C.~F.\ 1988, Reviews of Modern Physics, 60, 1 

\bibitem[Ptuskin \& Zirakashvili(2005)]{Ptuskin2005} Ptuskin, V.~S., \& Zirakashvili, V.~N.\ 2005, \aap, 429, 755 


\bibitem[Rho 
\& Borkowski(2002)]{Rho2002} Rho, J., \& Borkowski, K.~J.\ 2002, \apj, 575, 201 

\bibitem[Reach \& Rho(2000)]{Reach2000} Reach, W.~T., \& Rho, J.\ 2000, \apj, 544, 843

\bibitem[Reach et al.(2005)]{Reach2005} Reach, W.~T., Rho, J., \& Jarrett, T.~H.\ 2005, \apj, 618, 297 

\bibitem[Sano et al.(2010)]{Sano2010} Sano, H., Sato, J., Horachi, H., et al.\ 2010, \apj, 724, 59 

\bibitem[Smartt(2009)]{Smartt2009} Smartt, S.~J.\ 2009, \araa, 47, 63 

\bibitem[Tothill et al.(2002)]{Tothill2002} Tothill, N.~F.~H., White, G.~J., Matthews, H.~E., et al.\ 2002, \apj, 580, 285 


\bibitem[Vaupr{\'e} et al.(2014)]{Vaupre2014} Vaupr{\'e}, S., Hily-Blant, P., Ceccarelli, C., et al.\ 2014, \aap, 568, A50 

\bibitem[Vel{\'a}zquez et al.(2002)]{Velazquez2002} Vel{\'a}zquez, P.~F., Dubner, G.~M., Goss, W.~M., \& Green, A.~J.\ 2002, \aj, 124, 2145 

\bibitem[Voelk et al.(1984)]{Voelk1984} Voelk, H.~J., Drury, L.~O., \& McKenzie, J.~F.\ 1984, \aap, 130, 19


\bibitem[Tang(2017)]{Tang2017} Xiaping Tang\ 2017, arXiv:1707.00958

\bibitem[Yakovlev \& Pethick(2004)]{Yakovlev2004} Yakovlev, D.~G., \& Pethick, C.~J.\ 2004, \araa, 42, 169 

\bibitem[Yeung et al.(2016)]{Yeung2016} Yeung, P.~K.~H., Kong, A.~K.~H., Tam, P.~H.~T., et al.\ 2016, \apj, 827, 41

\bibitem[Yeung et al.(2017a)]{Yeung2017a} Yeung, P.~K.~H., Kong, A.~K.~H., Tam, P.~H.~T., et al.\ 2017, \apj, 837, 69

\bibitem[Zirakashvili 
\& Ptuskin(2008)]{Zirakashvili2008} Zirakashvili, V.~N., \& Ptuskin, V.~S.\ 2008, \apj, 678, 939

\bibitem[Zirakashvili 
\& Ptuskin(2012)]{Zirakashvili2012} Zirakashvili, V.~N., \& Ptuskin, V.~S.\ 2012, Astroparticle Physics, 39, 12 

\bibitem[Zirakashvili \& Ptuskin(2017)]{Zirakashvili2017} Zirakashvili, V.~N., \& Ptuskin, V.~S.\ 2017, arXiv:1701.00844

\bibitem[Zhou et al.(2014)]{Zhou2014} Zhou, P., Safi-Harb, S., Chen, Y., et al.\ 2014, \apj, 791, 87 

\end{thebibliography}
\end{document}